\documentclass[journal]{IEEEtran}
\usepackage{etex}
\reserveinserts{100}
\usepackage{mathtools}
\usepackage{graphics, theorem, times, amsfonts, graphicx, amssymb, cite}
\usepackage[outdir=figures_eps_to_pdf/]{epstopdf}
\usepackage{tikz}

\usetikzlibrary{shapes,arrows}
\usepackage{pgfplots}
\usepackage{color}
\usepackage{bbold}
\usepackage{setspace}
\usepackage{hyperref}
\usepackage{multirow}
\usepackage{rotating}
\usepackage{comment}
\usepackage{flushend}
\usepackage{caption}
\usepackage{subcaption}
\usepackage[affil-it]{authblk}
\usepackage{bm}
\usepackage{algorithm}
\usepackage{algorithmic}
\usepackage{enumerate}
\usepackage[shortlabels]{enumitem}
\usepackage{morefloats}
\usepackage{setspace, tikz}

\usetikzlibrary{matrix} 
\usetikzlibrary{arrows} 
\usetikzlibrary{calc} 
\usetikzlibrary{shapes}
\usetikzlibrary{arrows}
\usetikzlibrary{positioning}
\usetikzlibrary{backgrounds, fit}

\input{mysymbol.sty}
\usepackage{needspace}

\newcommand{\myparagraph}[1]{\needspace{1\baselineskip}\medskip\noindent {\bf #1.}}

\newcommand{\QED}{\hfill\ensuremath{\blacksquare}}

\def\E{\mathbb{E}}

\newtheorem{assumption}{\hspace{0pt}\bf Assumption}

\newtheorem{proposition}{\hspace{0pt}\bf Proposition}

\newtheorem{theorem}{\hspace{0pt}\bf Theorem}

\newtheorem{remark}{\hspace{0pt}\bf Remark}

\title{Optimal Wireless Resource Allocation with \\ Random Edge Graph Neural Networks}
\author{Mark Eisen$^{*}$ \qquad Alejandro Ribeiro$^{\dagger}$ 
\thanks{{Support by ARL DCIST CRA W911NF-17-2-0181 and Intel Science and Technology Center for Wireless Autonomous Systems. Authors are with $^{*}$Intel Corporation, Hillsboro, OR and $^{\dagger}$Univ. of Pennsylvania, Philadelphia, PA. Email: mark.eisen@intel.com, aribeiro@seas.upenn.edu}. Preliminary results presented in \cite{eisen2019large}.}
}

\begin{document}

\thispagestyle{empty}
\maketitle

\begin{abstract}
We consider the problem of optimally allocating resources across a set of transmitters and receivers in a wireless network. The resulting  optimization problem takes the form of constrained statistical learning, in which solutions can be found in a model-free manner by parameterizing the resource allocation policy. Convolutional neural networks architectures are an attractive option for parameterization, as their dimensionality is small and does not scale with network size. We introduce the random edge graph neural network (REGNN), which performs convolutions over random graphs formed by the fading interference patterns in the wireless network. The REGNN-based allocation policies are shown to retain an important permutation equivariance property that makes them amenable to transference to different networks. We further present an unsupervised model-free primal-dual learning algorithm to train the weights of the REGNN. Through numerical simulations, we demonstrate the strong performance REGNNs obtain relative to heuristic benchmarks and their transference capabilities. 
\end{abstract}


%

\section{Introduction} \label{sec_intro}

Wireless systems are integral to large scale intelligent systems, from robotics to the Internet of Things (IoT). The design of such systems requires optimal balancing of the numerous utilities and constraints that define the operating point of large networks of wireless connected devices. At a high level, such optimal design problems can be viewed as the allocation of a finite set of resources to achieve strong average performance over the randomly varying wireless channel. While these optimization problems can be easily formulated, they tend to be intractable as they are most often non-convex and infinite dimensional \cite{ribeiro2012optimal}. Some simplification is attained by working in the Lagrangian dual domain ~\cite{yu2006dual, ribeiro2012optimal} and subsequently using dual descent methods \cite{zhang2006stochastic, GatsisEtal15, wang2016dynamic}, or, alternatively, with heuristic optimization and scheduling methods \cite{shi2011iteratively, chen2011round, wu2013flashlinq, naderializadeh2014itlinq}. 

All such approaches invariably require accurate system models and may require prohibitively large computational cost. As emergent applications demand growth in scale and complexity, modern machine learning and statistical regression techniques have been explored as alternatives to solve wireless resource allocation problems. Machine learning methods train a learning model, such as a neural network (NN), to approximate the behavior of resource allocation strategies for a wide variety of problems. A common approach is to exploit supervised learning techniques to train a NN that approximates the behavior of an existing heuristic to reduce computational cost during execution \cite{sun2017learning, lei2017deep, xu2019energy, van2019sum}. The use of supervised learning, however, is limited by the availability of heuristics and hindered by their suboptimality. Supervised learning methods require these solutions to build training sets and may meet heuristic performance but never exceed it. 

Perhaps the more compelling justification for statistical learning solutions to wireless resource allocation problems lies in their reliance on \emph{data} over \emph{models}. That is, by treating the resource allocation problem itself as a form of statistical regression, we obtain a means of training policies that solve the optimal allocation problem directly rather than via a training set \cite{de2018team, xu2017deep, lee2018deep, eisen2019learning, liang2018towards, meng2019power}. This \emph{unsupervised} approach exceeds the capabilities of supervised learning in that it can be applied to any arbitrary resource allocation problem and has the potential to exceed performance of existing heuristics. In addition, it is possible to make unsupervised learning \emph{model-free} by relying on interactions with the wireless system. We probe with a candidate resource allocation policy, observe its outcome, and use this information to discover a better policy \cite{eisen2019learning}. 

There nonetheless remains the practical challenge of training models that can meet the scale of modern wireless systems. {\it Fully connected} neural networks (FCNNs) may seem appealing due to their well known universal approximation property \cite{sun2017learning, eisen2019learning}. However, FCNNs are also well known to be unworkable except in small scale problems. Scalability is attained in the processing of signals in time and space with {\it convolutional} neural networks (CNNs). Recognizing this fact has led to proposals that adapt CNNs to wireless resource allocation problems \cite{lee2018deep, van2019sum, xu2019energy}. A particularly enticing alternative is the use of a spatial CNN that exploits the spatial geometry of wireless networks to attain scalability to large scale systems with hundreds of nodes \cite{cui2019spatial}. 

In this paper we develop a different alternative to scalability that leverages graph neural networks (GNNs) \cite{henaff2015deep, gama2019convolutional}. GNNs are neural network architectures replacing the convolutional filter banks of CNNs with graph convolutional filter banks defined as polynomials on a matrix representation of a graph \cite{gama2019convolutional}. We propose here a variation which we call a random edge graph neural network (REGNN). REGNNs take as inputs the state of communication links and the state of the nodes of the network to produce resource allocation functions through the composition of layers which are themselves the composition of graph convolutional filter banks with pointwise nonlinearities. Through a combination of design choices as well as theoretical and numerical analyses this paper demonstrates that REGNNs have the following three properties:
\begin{description}
\item [Scalability.] REGNNs are defined by a number of parameters that is chosen independent of the number of nodes in the network. This enables training in large scale systems. We demonstrate in numerical experiments the possibility to scale to networks with several hundred nodes.
\item [Permutation invariance.] We prove that if REGNN parameters are optimal for a certain network, they are optimal for all of its permutations. This allows transference across different networks for as long as they are not far from permutations of each other. 
\item [Transference.] A given REGNN can be executed in any graph independent of shape and size. In particular, this makes it possible to train and execute in different networks. We demonstrate in numerical experiments the ability to transfer a REGNN across families of networks as well as the ability to train in networks of moderate size (with a few tens of nodes) and execute with good performance in large scale networks (with hundreds of nodes).
\end{description}
In conjunction with the model-free algorithmic learning approach developed in \cite{eisen2019learning}, we obtain a unified framework for learning effective resource allocation policies in large scale wireless systems. We point out related prior work \cite{eisen2019large, lee2019graph,shen2019graph} also considers graph neural network architectures for wireless resource allocation---see Remark \ref{remark_compare} for a comparison of these approaches.

We begin the paper by introducing a generic formulation of wireless resource allocation problems in which we seek a instantaneous resource allocation policy given a set of random fading states and random node states (Section \ref{sec_problem_formulation}). Such a formulation has many applications, ranging from multiple access to wireless control systems (Section \ref{sec_examples}). The problem generally cannot be solved exactly, but can be addressed through statistical learning techniques by parameterizing the resource allocation policy. We propose the use of random edge graph neural networks (REGNNs) to parameterize the policy by viewing the random fading links between transceivers as a graph with random edges (Section \ref{sec_gnn}). This parameterization is a generalization of the popular convolution neural networks and has low-dimensionality that makes it scalable for large wireless networks.

Any policy of practical use should have the ability to be implemented on varying network topolgies. We present in Theorem \ref{theo_invariance} a so-called permutation invariance of the REGNN with respect to the underlying graph structure of its inputs (Section \ref{sec_equivariance}). We further establish the permutation equivariance of both the optimal unparameterized resource allocation policy (Section \ref{sec_perm}) and the learned REGNN (Section \ref{sec_equivariance_regnns}). We present an unsupervised, model-free primal-dual algorithm to train the REGNN filter tensor weights without requiring explicit model knowledge (Section \ref{sec_primal_dual}). We conclude with a comprehensive set of numerical simulations that demonstrate the strong performance of learned REGNN resource allocation policies (Section \ref{sec_numerical_results}), including their ability to transfer to varying network topologies (Section \ref{sec_transference}). 

%
%
%

%

%
\begin{figure*}
\centering
\begin{subfigure}[b]{0.47\linewidth}
   \centering
   \includegraphics[width=\linewidth]{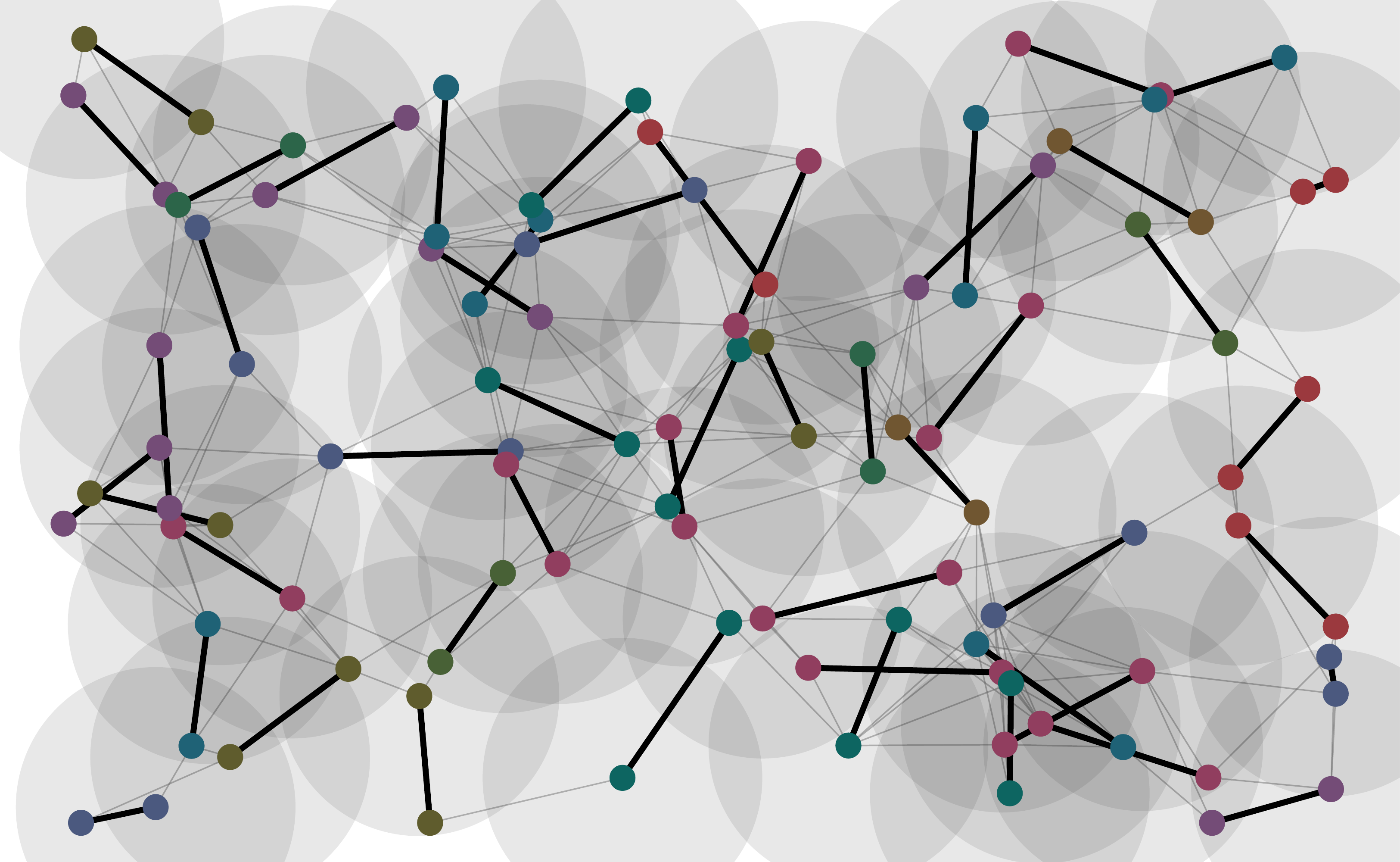}        
   \caption{Wireless ad-hoc network}
   \label{fig_systema}
\end{subfigure} 
\qquad
\begin{subfigure}[b]{0.47\linewidth}
   \centering
   \includegraphics[width=\linewidth]{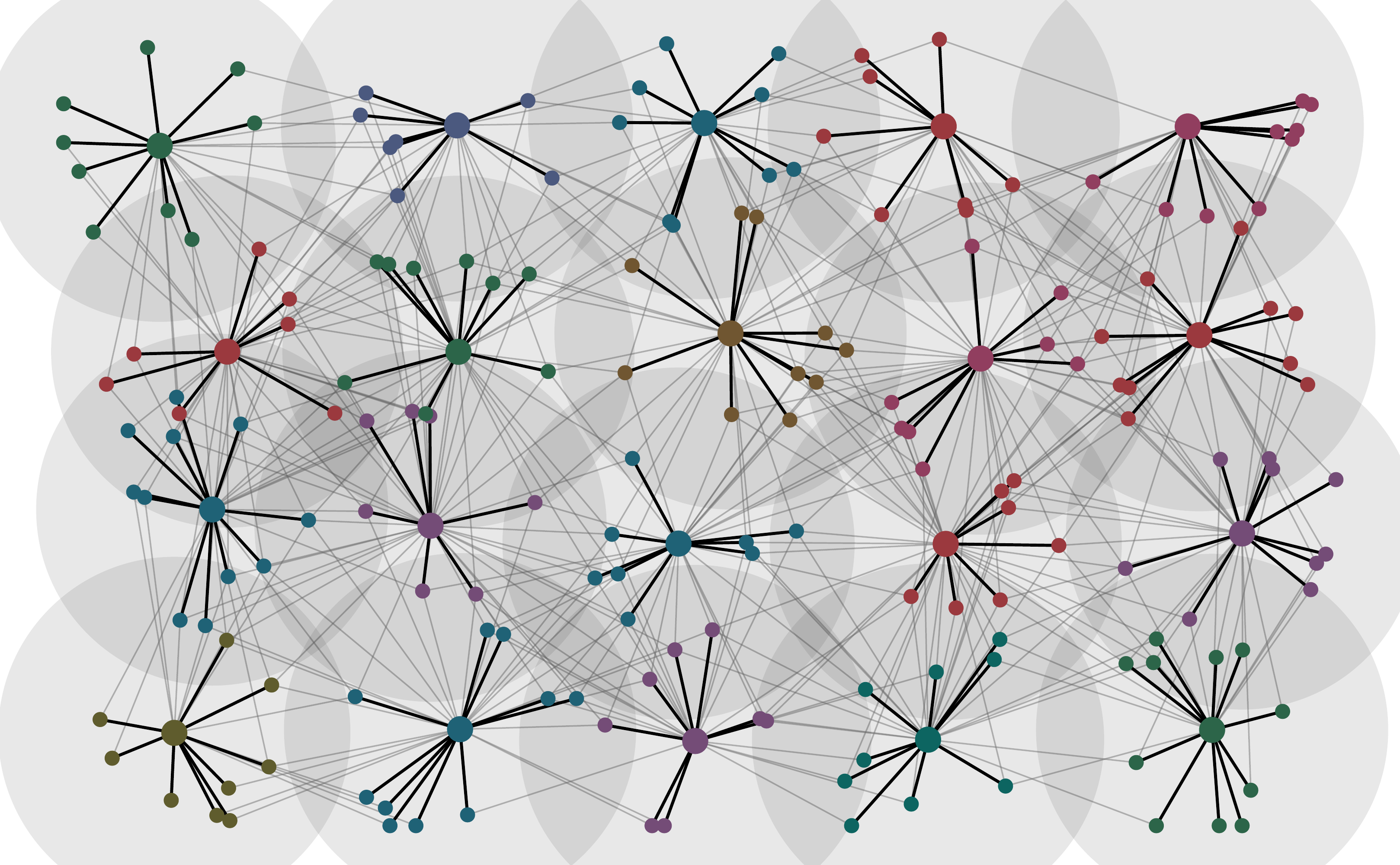}
   \caption{Cellular communication uplink}
   \label{fig_systemb}
\end{subfigure}
\caption{Resource allocation in large scale wireless networks. Wireless links connect transmitters to receivers (bold lines) but wireless transmission also generates interference to other receivers (thin lines). We formulate optimal resource allocation as machine learning problem over the interference graph that we solve using random edge graph neural networks (REGNNs).}
\label{fig_system}
\end{figure*}

%
\section{Optimal Resource Allocation in Wireless Communication Systems} \label{sec_problem_formulation}

Consider a wireless system made up of a set of $m$ transmitters and $n$ receivers. Each transmitter $i\in\{1,m\}$ is paired with a single receiver $r(i)\in\{1,n\}$. Multiple transmitters may be paired with the same receiver. We denote as $\ccalR_j : =\{i: r(i) = j\}$ the set of transmitters $i$ paired with receiver $j$. Figure \ref{fig_systema} illustrate ad hoc networks in which $n=m$ and the pairing of transmitters and receivers is bijective. Figures \ref{fig_systemb} illustrates a cellular uplink in which $m>n$ and the map is surjective with $\ccalR_j$ containing the users in the catchment area of base station $j$. To model a cellular downlink we need to replicate the base station node to produce a bijective map from base stations to nodes. Although not required we have in mind large scale wireless systems with $n$ ranging from several tens to several hundreds.

Time is slotted with connections between agents in a time slot characterized by fading channel coefficients. We use $h_{ii}(t)$ to denote the channel between transmitter $i$ and receiver $r(i)$ and $h_{ij}(t)$ to denote the channel between transmitter $i$ and receiver $r(j)$ at time slot $t$. All channels are arranged in the matrix $\bbH(t)\in\reals^{m\times m}$ with entries $[\bbH(t)]_{ij} = h_{ij}(t)$. In addition to channel states there are also separate state variables $x_i(t)$ representing a random state of the communication between $i$ and $r(i)$ -- such as, e.g., the number of packets that arrive in the time slot. These node states are collected in the vector $\bbx(t)$ with entries $[\bbx(t)]_i = x_i(t)$. Our goal is to map state observations $\bbH(t)$ and $\bbx(t)$ to a resource allocation function $\bbp(t) = \bbp(\bbH(t),\bbx(t))$. Allocating resources $\bbp(t)$ when the system state pair is $\bbH(t), \bbx(t)$ produces a vector reward of the form
\begin{align}\label{eqn_instantaneous_reward}
   \bbr(t) = \bbf\Big(\bbp\big(t\big); \bbH(t), \bbx(t)\Big),
\end{align}
where $\bbf$ is a function from the joint space of resource allocations and states to the space of rewards. This abstract model encompasses several problems of practical importance as we illustrate in Section \ref{sec_examples}.

In fast fading scenarios the instantaneous value of the reward $\bbr(t)$ in \eqref{eqn_instantaneous_reward} is not directly experienced by end users. Rather, end users experience the long term average across time slots. Assuming stationarity and independence of subsequent state realizations, this time average can be replaced by an expectation. Let then $m(\bbH, \bbx)$ represent a probability distribution of channel and node states and approximate the long term average reward by its limit which we can equate to the expected reward, 
\begin{align}\label{eqn_average_reward}
   \bbr\! = \mbE\Big[ \bbf\Big(\bbp\big(\bbH,\bbx\big); \bbH, \bbx\Big) \Big]
       \! =  \int\!\bbf\Big(\bbp\big(\bbH,\bbx\big); \bbH, \bbx\Big)  d m(\bbH, \bbx).
\end{align}
The goal of the optimal wireless system design problems we study in this paper is to find the instantaneous resource allocation policy $\bbp(\bbH, \bbx)$ that optimizes the expected reward in \eqref{eqn_average_reward}. Specifically, introduce a utility function $u_0(\bbr)$ and a set of utility constraints $\bbu(\bbr)\leq \bb0$ to formulate the optimization problem
\begin{alignat}{3} \label{eq_problem}
   \bbp^*(\bbH,\bbx) 
       \ = \ & \argmax\quad && u_0(\bbr),                                                            \\ \nonumber 
             & \st          && \bbr = \mbE\Big[\bbf\Big(\bbp\big(\bbH,\bbx\big);\bbH,\bbx\Big)\Big], \\ \nonumber 
             &              && \bbu(\bbr) \geq \bb0, \quad 
                               \bbp(\bbH,\bbx) \in \ccalP (\bbH,\bbx). 
\end{alignat}
In the problem in \eqref{eq_problem} we find an average reward $\bbr$ that maximizes the utility $u_0(\bbr)$ while making sure the utility constraints $\bbu(\bbr)\leq \bb0$ are satisfied. We do so by searching for the resource allocation $\bbp^*(\bbH,\bbx)$ that produces such expected reward according to \eqref{eqn_average_reward}. We have also added the constraint $\bbp(\bbH,\bbx) \in \ccalP (\bbH,\bbx)$ to represent (simple) constraints on allowable resource allocations; see, Section \ref{sec_examples}.

Note that most prior heuristic resource allocation methods maximize an instantaneous reward instead of the average reward considered in \eqref{eq_problem}. Generally speaking, maximizing an instantaneous reward requires the solving of a constrained optimization problem for each instance of the channel and node states $\bbH$ and $\bbx$---for complex problems with fast varying channel conditions, it often poses an infeasible computational burden to solve the problem for every channel instance. The formulation in \eqref{eq_problem} defines a single policy $\bbp(\bbH,\bbx)$ that maximizes the ergodic performance of the system; thus, the global optimization need only be solved once offline. At execution, the instantaneous resource allocation is obtained by passing instantaneous states through $\bbp(\bbH,\bbx)$ at marginal computational cost \cite{ribeiro2012optimal}.

The utilities in \eqref{eq_problem} are design choices and can be made convex. The ergodic constraint [cf. \eqref{eqn_average_reward}], however, incorporates the function $\bbf$ in \eqref{eqn_instantaneous_reward} which is typically not convex. In addition, fading channel realizations are a dense set and we are interested in cases where the number of transmitters is large. This makes solution of \eqref{eq_problem} intractable and motivates the use of various heuristics; e.g., \cite{shi2011iteratively}. If we are considering heuristics in general, we can, in particular, use data driven heuristics where we propose some resource allocation, observe its outcome, and use this information to update the resource allocation policy. To that end we follow an interpretation of \eqref{eq_problem} originally developed in \cite{eisen2019learning}, in which we identify it as a constrained statistical learning problem. Consider then a parameter $\bbtheta\in\reals^q$ and a function family $\bbPhi\big(\bbH,\bbx;\bbtheta\big)$ that we use to generate resource allocations
\begin{align} \label{eq_learning_parametrization}
  \bbp\big(\bbH,\bbx\big) = \bbPhi\big(\bbH,\bbx;\bbtheta\big) .
\end{align}
To fix ideas, say that we choose the family $\bbPhi$ to consist of quadratic plus linear functions of the form $\bbPhi(\bbH,\bbx, \bbtheta) = (1/2)\bbtheta_1^T\bbH\bbtheta_1 + \bbtheta_2^T \bbx$. Or, as done in many recent machine learning applications, we make $\bbPhi(\bbH,\bbx, \bbtheta)$ the output of a neural network; e.g., \cite{sun2017learning, liang2018towards, eisen2019learning}. In any event, with a given parametrization we can substitute \eqref{eq_learning_parametrization} into \eqref{eq_problem} to obtain a problem in which the optimization over resource allocations $\bbp(\bbH,\bbx)$ is replaced with an optimization over the set of parameter vectors $\bbtheta$,
\begin{alignat}{3} \label{eq_param_problem}
   \bbtheta^* \ = \ & \argmax\ && u_0(\bbr),           \nonumber \\
                & \st      && \bbr = \mbE\Big[ \bbf\Big(\bbPhi\big(\bbH,\bbx;\bbtheta\big); \bbH, \bbx\Big) \Big], \nonumber \\ 
                &          && \bbu(\bbr) \geq \bb0, \quad 
                               \bbPhi(\bbH,\bbx; \bbtheta) \in \ccalP (\bbH,\bbx).
\end{alignat}
Notice that in \eqref{eq_param_problem} we include the hard constraint $\bbPhi(\bbH,\bbx; \bbtheta) \in \ccalP (\bbH,\bbx)$. We point out that this constraint can often be satisfied in practice through careful design of $\bbPhi(\bbH,\bbx; \bbtheta)$. In particular, the parameterization can be augmented to include an additional output operation that projects the allocation onto $\ccalP(\bbH,\bbx)$.

The problems in \eqref{eq_problem} and \eqref{eq_param_problem} look similar but are different in three important ways: (i) The original optimization problem in \eqref{eq_problem} is a functional optimization problem given that, in general, $\bbH$ and $\bbx$ belong to dense sets. The parametrized problem in \eqref{eq_param_problem} is on the $q$-dimensional variable $\bbtheta$. (ii) The parametrized problem in \eqref{eq_param_problem} can be solved without having access to a model for the function $\bbf$. It suffices to have the ability to probe the system with a resource allocation $\bbPhi(\bbH,\bbx;\bbtheta)$ and measure the outcome $\bbf(\bbPhi(\bbH,\bbx;\bbtheta);\bbH,\bbx)]$ as we detail in Section \ref{sec_primal_dual} -- see also \cite{eisen2019learning}. This is impossible in \eqref{eq_problem} whose solution requires access to the model $\bbf$. (iii) The learning parametrization reduces the space of allowable resource allocations so that the optimal solution of \eqref{eq_param_problem} entails a loss of optimality relative to the solution of \eqref{eq_problem}. 

The latter point calls for judicious choice of the learning parametrization. E.g., if we use a fully connected neural network in \eqref{eq_learning_parametrization} we can rely on universality results to claim a small loss of optimality -- along with other interesting theoretical claims \cite{eisen2019learning}. However, fully connected neural networks do not work beyond simple low dimensional problems and our interest is in problems where we have $m^2$ input variables in $\bbH$ and $m$ input variables in $\bbx$ with large $m$. We propose here to use graph neural networks (Section \ref{sec_gnn}) which we will demonstrate provide an scalable parametrization that permits finding good solutions to \eqref{eq_param_problem} with large values of $m$ (Section \ref{sec_numerical_results}). Before introducing GNNs we present some examples of resource allocation problems belonging to the family of abstract problems introduced in \eqref{eq_problem}.

%
\subsection{Examples}\label{sec_examples}

%
\myparagraph{Multiple access AWGN channel} Terminals communicating with associated receivers on a shared channel. A standard instantaneous performance metric of interest here is the capacity experienced by each user under noise and interference. The $i$th element of $\bbf(\bbp(\bbH,\bbx; \bbH, \bbx)$ may then denote the instantaneous capacity achieved by transmitter $i$. In a channel subject to additive white Gaussian noise (AWGN) and multi-user interference and assuming the use of capacity achieving codes this is given by 
\begin{align}\label{eq_interference_problem1}
  f_i(\bbp; \bbH, \bbx) :=  \log \bigg(1 + \frac{h_{ii} p_i(\bbH,\bbx)}
              {1 + \sum_{j \neq i} h_{ji} p_j(\bbH,\bbx)}\bigg).
\end{align}
Defining the performance as in \eqref{eq_interference_problem1} in the constraint in \eqref{eq_problem} reflects a maximization with respect to the long term capacity experienced by the users. A constraint of the form $\bbu(\bbr)\geq\bb0$ enforces a  minimum average capacity for all users if $\bbu(\bbr) := \bbr - \bbc_{\min}$ to achieve fairness. Power constraints can be enforced via the set $\ccalP = \{\bbp: \bb0 \leq \bbp \leq \bbp_0\}$ and the utility $u_0$ can be chosen to be the sum rate $u_0(\bbr) = \sum_{i} r_i$ or a proportional fair utility  $u_0(\bbr) = \sum_{i} \log(r_i)$. In this problem there is no node state $\bbx$ that plays a role in the system's design. 

%
\myparagraph{Multiple access with user demand} We augment the previous example to incorporate varying traffic demand or information generation at each node. Here, the state $x_i$ reflects the rate of information collection or the data arrival rate at the $i$th transmitter. An additional performance metric is the average arrival rate $\bbr^\prime := \E (\bbx)$. A necessary constraint on the system is that average capacity exceeds observed or target average collection rates $\bbr_0$. We achieve that with the constraint
\begin{align}\label{eq_data_problem}
  \bbu(\bbr^{\prime}) := \bbr_0 - \bbr^{\prime} \geq \bb0.
\end{align}
Another form of user demand comes in the context of fairness, in which the state $\bbx$ may reflect the achieved lifetime rates of all of the users with ergodic average $\bby := \bbE (\bbx)$. A weighted sum-rate would thus prioritize users who have not been given sufficient access to the channel, i.e.,
\begin{align}\label{eq_weighted_Rate}
\bbu(\bbr,\bby) = \sum_i \frac{r_i}{y_i}.
\end{align}
These problems make full use of the generality of \eqref{eq_problem} by containing both the fading channel states and associated performance metric in the capacity function given in \eqref{eq_interference_problem1}, as well as a node state given by either data collection and associated coupled constraint given by \eqref{eq_data_problem}, or lifetime rate and associated weighted objective given by \eqref{eq_weighted_Rate}. 

%
\myparagraph{Random access wireless control systems} A more complex example modeled by \eqref{eq_problem} concerns resource allocation in a wireless control system. Consider that transmitters are sending plant state information to a shared receiver/base station to compute control inputs over a common random access channel that is subject to potential packet collisions. Given the direct and interference channel states and transmission powers, we define a function $q(p_i, h_{ii}, p_j,h_{ij}) \rightarrow [0,1]$ that gives a probability of collision between transmitters $i$ and $j$. We are interested in the probability of successful transmission of transmitter $i$, i.e.,
\begin{align}\label{eq_interference_problem2}
  q_i(\bbp; \bbH, \bbx) := \prod_{j \neq i} \left(1 - q(p_i,h_{ii}, p_j, h_{ij}) \right).
\end{align}
Likewise, consider the node state $x_i$ to be the state of the plant at the $i$th transmitter. If transmission is succesful, the system state evolves with stable gain $1 > \gamma_c > 0$; otherwise, it evolves with unstable gain gain $\gamma_o > 1$. We are often concerned with a quadratic cost that measures the one step distance from the origin of the plant state, which can be written as the following cost (or negative reward), i.e.
\begin{align}\label{eq_control_cost}
f_i(\bbp; \bbH, \bbx) := q_i(\bbp; \bbH, \bbx)(\gamma_c x_i)^2 + (1-q_i(\bbp; \bbH, \bbx))(\gamma_o x_i)^2.
\end{align}
The cost metric \eqref{eq_control_cost} can be used to define a utility that maximizes the expected negative cost, i. e. $u_0(\bbr) := - \mathbf{1}^T \bbr $ and constraints that impose a minimum long-term cost $\kappa_{\max}$ for each plant, i. e. $u_i(r_t) := r_i - \kappa_{\max} \leq 0 $.

%

%
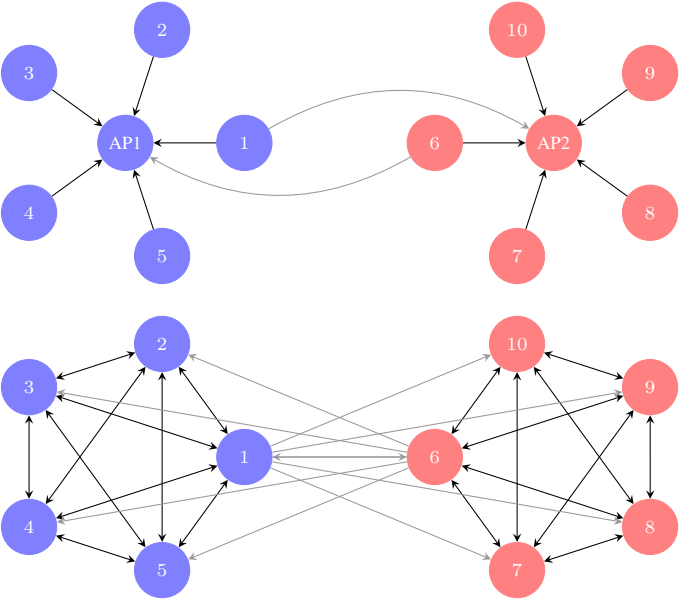
\begin{figure}
\centering

\tikzstyle{empty node} = [ circle, 
                           draw = black,
                           text = black, 
                           minimum size = 0.65*\unit]

\tikzstyle{blue node} = [ empty node, 
                         fill = blue!50,
                         draw = blue!50,
                         text = white]

\tikzstyle{red node} = [ empty node, 
                         fill = red!50,
                         draw = red!50,
                         text = white]

\tikzstyle{edge} = [shorten >=0pt, shorten <=0pt]

\def \myfactor {1.13}
\def \unit     {\myfactor cm}

\def \radius   {1.4}

{\fontsize{7}{7}\selectfont\begin{tikzpicture}[scale = \myfactor]

  \node                             [blue node] (AP1) {AP1}; 
  \path (AP1) ++ (  0:\radius) node [blue node] (1)   {$1$}; 
  \path (AP1) ++ ( 72:\radius) node [blue node] (2)   {$2$}; 
  \path (AP1) ++ (144:\radius) node [blue node] (3)   {$3$}; 
  \path (AP1) ++ (216:\radius) node [blue node] (4)   {$4$}; 
  \path (AP1) ++ (288:\radius) node [blue node] (5)   {$5$}; 

  \path (AP1) ++ (3.6*\radius,0) node [red node] (AP2) {AP2};  
  \path (AP2) ++ (180:\radius)   node [red node] (6)   {$6$};  
  \path (AP2) ++ (252:\radius)   node [red node] (7)   {$7$};  
  \path (AP2) ++ (324:\radius)   node [red node] (8)   {$8$};  
  \path (AP2) ++ ( 36:\radius)   node [red node] (9)   {$9$}; 
  \path (AP2) ++ (108:\radius)   node [red node] (10)  {$10$}; 

  \path[-stealth] (1)  edge [edge] node {} (AP1);
  \path[-stealth] (2)  edge [edge] node {} (AP1);
  \path[-stealth] (3)  edge [edge] node {} (AP1);
  \path[-stealth] (4)  edge [edge] node {} (AP1);
  \path[-stealth] (5)  edge [edge] node {} (AP1);

  \path[-stealth] (6)  edge [edge] node {} (AP2);
  \path[-stealth] (7)  edge [edge] node {} (AP2);
  \path[-stealth] (8)  edge [edge] node {} (AP2);
  \path[-stealth] (9)  edge [edge] node {} (AP2);
  \path[-stealth] (10) edge [edge] node {} (AP2);
  
  \path[black!40, -stealth] (1)  edge [edge, bend left] node {}  (AP2);

  \path[black!40, -stealth] (6)  edge [edge, bend left] node {} (AP1);

\end{tikzpicture}}

\bigskip

{\fontsize{7}{7}\selectfont\begin{tikzpicture}[scale = \myfactor]

  \node                             []          (AP1) {};    
  \path (AP1) ++ (  0:\radius) node [blue node] (1)   {$1$}; 
  \path (AP1) ++ ( 72:\radius) node [blue node] (2)   {$2$}; 
  \path (AP1) ++ (144:\radius) node [blue node] (3)   {$3$}; 
  \path (AP1) ++ (216:\radius) node [blue node] (4)   {$4$}; 
  \path (AP1) ++ (288:\radius) node [blue node] (5)   {$5$}; 

  \path (AP1) ++ (3.6*\radius,0) node []         (AP2) {};     
  \path (AP2) ++ (180:\radius)   node [red node] (6)   {$6$};  
  \path (AP2) ++ (252:\radius)   node [red node] (7)   {$7$};  
  \path (AP2) ++ (324:\radius)   node [red node] (8)   {$8$};  
  \path (AP2) ++ ( 36:\radius)   node [red node] (9)   {$9$}; 
  \path (AP2) ++ (108:\radius)   node [red node] (10)  {$10$}; 

  \path[stealth-stealth] (1)  edge [edge] node {} (2);
  \path[stealth-stealth] (1)  edge [edge] node {} (3);
  \path[stealth-stealth] (1)  edge [edge] node {} (4);
  \path[stealth-stealth] (1)  edge [edge] node {} (5);
  \path[stealth-stealth] (2)  edge [edge] node {} (3);
  \path[stealth-stealth] (2)  edge [edge] node {} (4);
  \path[stealth-stealth] (2)  edge [edge] node {} (5);
  \path[stealth-stealth] (3)  edge [edge] node {} (4);
  \path[stealth-stealth] (3)  edge [edge] node {} (5);
  \path[stealth-stealth] (4)  edge [edge] node {} (5);

  \path[stealth-stealth] (6)  edge [edge] node {}  (7);
  \path[stealth-stealth] (6)  edge [edge] node {}  (8);
  \path[stealth-stealth] (6)  edge [edge] node {}  (9);
  \path[stealth-stealth] (6)  edge [edge] node {} (10);
  \path[stealth-stealth] (7)  edge [edge] node {}  (8);
  \path[stealth-stealth] (7)  edge [edge] node {}  (9);
  \path[stealth-stealth] (7)  edge [edge] node {} (10);
  \path[stealth-stealth] (8)  edge [edge] node {}  (9);
  \path[stealth-stealth] (8)  edge [edge] node {} (10);
  \path[stealth-stealth] (9)  edge [edge] node {} (10);

  \path[black!40, -stealth] (1)  edge [edge] node {}  (6);
  \path[black!40, -stealth] (1)  edge [edge] node {}  (7);
  \path[black!40, -stealth] (1)  edge [edge] node {}  (8);
  \path[black!40, -stealth] (1)  edge [edge] node {}  (9);
  \path[black!40, -stealth] (1)  edge [edge] node {}  (10);

  \path[black!40, -stealth] (6)  edge [edge] node {}  (1);
  \path[black!40, -stealth] (6)  edge [edge] node {}  (2);
  \path[black!40, -stealth] (6)  edge [edge] node {}  (3);
  \path[black!40, -stealth] (6)  edge [edge] node {}  (4);
  \path[black!40, -stealth] (6)  edge [edge] node {}  (5);

\end{tikzpicture}}
\caption{Communication network and interference graph. Random edge graph neural networks (Section \ref{sec_gnn}) run on the interference graph (bottom), not the communication graph (top). Nodes 1-5 communicate with AP1 and nodes 6-10 with AP2 while node 1 also interferes on AP2 and node 6 on AP1. In the interference graph nodes that communicate or interfere with an AP form a clique. The APs are not nodes of the interference graph.}
\label{fig_g_signal}
\end{figure}

%
\section{Random Edge Graph Neural Networks}\label{sec_gnn}

For the learning parametrization in \eqref{eq_param_problem} we introduce random edge (RE) graph neural networks (GNNs). To that end, recall the definition of $\bbx = [x_1; \ldots; x_m]$ as a vector whose entry $x_i$ represents the state of the communication from node $i$ to receiver $r(i)$ and reinterpret $\bbx$ as a signal supported on the nodes $i=1,\hdots,m$. Further reinterpret the channel matrix $\bbH \in \reals^{m \times m}$ as an adjacency matrix representation of a graph linking node $i$ to node $j$. REGNNs rely on graph convolutional filters supported on the graph $\bbH$ to process some input signal $\bbz \in \reals^{m}$. Formally, let $\bbalpha := [\alpha_0; \ldots; \alpha_{K-1}]$ be a set of $K$ filter coefficients and define the graph filter $\bbA(\bbH)$ as a polynomial on the graph representation that is linearly applied to an input signal $\bbz $ to produce the output signal,
\begin{align}\label{eq_graph_conv}
   \bby \  = \ \bbA(\bbH) \bbz 
        \ := \ \sum_{k=0}^{K-1} \alpha_k \bbH^k \bbz.
\end{align}
In the graph signal processing literature, the filter $\bbA(\bbH) = \sum_{k=0}^{K-1} \alpha_k \bbH^k$ is said to be a linear shift invariant filter and the matrix $\bbH$ is a graph shift operator (GSO)\cite{sandryhaila2014big}. If we particularize $\bbH$ to represent a cyclic graph, the operation in \eqref{eq_graph_conv} reduces to the conventional convolution operation. 

For an intuitive understanding of the graph filter in \eqref{eq_graph_conv}, consider that a single shift operation $\bbH \bbz$ will aggregate information at each node from its immediate neighbors scaled by their associated edge weights; likewise, a $k$-shift operation $\bbH^k \bbz$ will aggregate information from the $k$-hop neighborhood. This can be considered as a multi-hop message passing of local information between nodes. A benefit of larger graph filters in a GNN is that each set of node features $\bbz$ will be a generated from a more complete picture of the network. For instance, a graph filter of length $K=2$ will generate a feature using the fading state information of a node and its neighbors, but will not directly consider the fading state information of the more distant nodes in the network---see Remark \ref{rmk_locality} for a discussion of this locality in regards to the proposed architecture.

To define a REGNN we compose $L$ layers, each of which is itself the composition of a graph filter with a pointwise nonlinearity. Introduce then a layer index $l$ in \eqref{eq_graph_conv} so that we have $K_l$ filter coefficients 
$\bbalpha_l = [\alpha_{l0}; \ldots; \alpha_{l(K_l-1)}]$ defining graph filters $\bbA_l(\bbH) = \sum_{k=0}^{K_l-1} \alpha_{lk} \bbH^k$. We apply the filter $\bbA_l(\bbH)$ to the output of layer $l-1$ to produce the layer $l$ intermediate feature $\bby_l   =  \bbA(\bbH) \bbz_{l-1}$. This intermediate feature is then passed through a pointwise nonlinearity function  $\bbsigma:\reals^m\to\reals^m$  to produce the output of the $l$th layer as
\begin{align}\label{eq_single_feature_gnn}
   \bbz_l = \bbsigma \big [ \bby_{l}                 \big ]
          = \bbsigma \Big [ \bbA_l(\bbH) \bbz_{l-1}  \Big ]
          = \bbsigma \Bigg[ \sum_{k=0}^{K_l-1} \alpha_{lk} \bbH^k  \bbz_{l-1}  \Bigg]   
\end{align}
A REGNN is defined by recursive application of \eqref{eq_single_feature_gnn}. The input to this recursion is the node state vector, i.e. $\bbz_0 := \bbx$. The output is the $l$th layer signal $\bbz_L$. We emphasize that the nonlinear function $\bbsigma$ in \eqref{eq_single_feature_gnn} is applied individually to each component. Namely, for any input vector $\bbv$ we must have $[\bbsigma(\bbv)]_i = \bbsigma([\bbv]_i)$. Common choices for $\bbsigma$ are rectified linear units (ReLu), absolute values, or sigmoids \cite{henaff2015deep, gama2019convolutional}. 

To increase the expressive power of REGNNs we consider multiple features per layer. That is, instead of processing the output of layer $l-1$ with a single graph filter, we process it with a bank of $F_l$ graph filters. This process generates multiple features per layer, each of which we process with a separate graph filter bank. Suppose then that the output of layer $l-1$ consists of $F_{l-1}$ features $\bbz_{l}^f$. These features become inputs to layer $l$, each of which we process with a $F_l$ filters $\bbA_l^{fg}(\bbH)$ defined by coefficients $\bbalpha_{l}^{fg} := [\alpha^{fg}_{l0}; \ldots; \alpha^{fg}_{l(K_l-1)}]$. Applying each of these filters to each of the input features produce the $l$th layer intermediate features
\begin{align}\label{eqn_gnn_intermediate_features}
   \bby_{l}^{fg} \  = \ \bbA_l^{fg}(\bbH) \bbz_{l-1}^f 
                 \  = \ \sum_{k=0}^{K_l-1} \alpha_{lk}^{fg} \bbH^k \bbz_{l-1}^f .
\end{align}
The $l$th layer filter bank therefore produces a total of $F_{l-1} \times F_l$ intermediate features $\bby_{l}^{fg}$. To avoid exponential growth of the number of features all features $\bby_{l}^{fg}$ for a given $g$ are linearly aggregated and passed through the pointwise nonlinearity $\bbsigma$ to produce the $l$th layer output
\begin{align}\label{eq_layer_l_output}
   \bbz_{l} = \bbsigma_l \Bigg [  \sum_{f=1}^{F_l} \bby_{l}^{fg}                  \Bigg]
            = \bbsigma_l \Bigg [  \sum_{f=1}^{F_l} \bbA_l^{fg}(\bbH) \bbz_{l-1}^f \Bigg].
\end{align}
The REGNNs we consider in this paper are defined by recursive application of \eqref{eq_layer_l_output}. The input to layer $l=1$ is the (single feature) signal $\bbx = \bbz_0^1$. The output of the REGNN is the (also single feature)  signal $\bbz_{L} = \bbz_L^1$. For future reference we group all filter coefficients in the filter tensor $\bbA = \{\alpha_{lk}^{fg}\}_{l,f,g,k}$ and define the REGNN operator as,
\begin{align}\label{eq_gnn_param}
   \bbPhi (\bbH, \bbx; \bbA) = \bbz_L.
\end{align}
The operator in \eqref{eq_gnn_param} is a graph neural network \cite{henaff2015deep, gama2019convolutional} if we fix the graph $\bbH$. Here we call it a random edge (RE)GNN to emphasize that $\bbH$ is an input to the operator $\bbPhi$.

Our goal is to solve \eqref{eq_problem} using as inputs the class of functions that can be represented by the REGNNs in \eqref{eq_gnn_param}. This translates to solving the optimization problem
\begin{alignat}{3} \label{eq_gnn_param_problem}
   \bbA^* \ = \ & \argmax\ && u_0(\bbr),           \nonumber \\
                & \st      && \bbr = \mbE\Big[ \bbf\Big(\bbPhi\big(\bbH,\bbx;\bbA\big); \bbH, \bbx\Big) \Big], \nonumber \\ 
                &          && \bbu(\bbr) \geq \bb0,  \quad 
                               \bbPhi(\bbH,\bbx; \bbA) \in \ccalP (\bbH,\bbx).  
\end{alignat}
The REGNN receives as input a random graph signal input $\bbx \sim m(\bbx)$ and a random underlying graph shift operator $\bbH \sim m(\bbH)$. According to \eqref{eq_gnn_param_problem}, the filter coefficients in the tensor $\bbA$ are trained relative to the statistics of both of these quantities; the graph signal $\bbx$ and the underlying graph $\bbH$. 

Notice that to solve \eqref{eq_gnn_param_problem} we need to specify the class of admissible GNN representations $\bbPhi (\bbH, \bbx; \bbA)$. This requires specifying the number of layers, $L$, the number of features at each layer, $F_l$, and the length of the filters used at each layer, $K_l$. We say that $L$, $F_l$, and $K_l$ specify the REGNN architecture. The total number of parameters that specify this architecture is $q = \sum_{l=1}^L  \sum_{l=1}^{L} K_l \times F_l \times F_{l+1}$, which simplifies to $LKF^2$ if all layers use $F_l=F$ features and filters of length $K_l=K$. This number is (much) smaller than the number of parameters that would be required to train a fully connected neural network. We further point out that the feasibility condition $\bbPhi(\bbH,\bbx; \bbA) \in \ccalP (\bbH,\bbx)$ can often be easily addressed by utilizing the final output layer activation $\bbsigma_L$ to project onto $\ccalP(\bbH,\bbx)$.

Perhaps most importantly, and as will be seen in the following section, the use of graph filters creates a permutation equivariance that matches the permutation equivariance of the optimal solution of \eqref{eq_problem}. This equivariance suggests that \mbox{REGNNs} likely generalize across different network realizations; something we will verify in the numerical experiments in Section \ref{sec_numerical_results}.

%
\begin{remark}[Graph shift operator]\label{rmk_graph}\normalfont 
The graph $\bbH$ is an asymmetric graph with self loops in which the weight in the edge $(i,j)$ is the fading channel realization $h_{ij}$. Notice that this edge weight is {\it not} the strength of the channel linking node $i$ to node $j$ but the strength of the channel linking transmitter $i$ to the receiver $r(j)$ associated to node $j$. Further observe that this is a random graph whose realizations are drawn from the distribution $m(\bbH)$. The mapping of interference patterns to the graph structure can be visualized in Figure \ref{fig_g_signal}.
\end{remark}

%
\begin{remark}[Locality of Graph Filters]\label{rmk_locality}\normalfont Observe in \eqref{eq_graph_conv} that the $k$th filter tap scales the input by the $k$th order of the graph shift $\bbH$. This term reflects a $k$-hop shift of the elements in $\bbz$, with each hop weighted by the associated edge. As the order $k$ increases, node states from larger neighborhoods are incorporated. Thus, the \emph{locality} of a node, or the weight of its $k$ hop neighborhood, provides a guide for selecting filter size $K$, as incorporating higher order information will have diminishing impact as the $k$th order neighborhood shrinks in size or weight.
\end{remark}

%
\begin{remark}[Convolutional Neural Networks]\label{rmk_cnns}\normalfont Just as the standard convolution operation is a particular case of the graph filter in \eqref{eq_graph_conv} for the cyclic graph, the REGNN generalizes the standard   convolutional neural network (CNN) to include random and arbitrary graph structures. CNNs have been empirically observed to be strikingly effective in many learning tasks ranging from image classification \cite{krizhevsky2012imagenet} to recommender systems \cite{cheng2016wide}. Their success is not attributed only to their low dimensionality, but by the fact that they contain a translation equivariance property necessary for, e.g., image classification. As the REGNN is a generalization of the CNN, it is reasonable to expect that they contain similar equivariances---namely, an invariance to permutations. In Section \ref{sec_perm} we establish a permutation equivariance of optimal wireless resource allocation, and proceed to establish the same property in REGNNs.
\end{remark}

%
\begin{remark}[Related Work]\label{remark_compare}\normalfont 
We point out the related prior work that utilize graph neural network structures for wireless resource allocation problems \cite{eisen2019large, lee2019graph,shen2019graph}. Our preliminary work in \cite{eisen2019large} introduces the notion of using random fading states to represent a graph in a graph neural network. The works in \cite{lee2019graph,shen2019graph} also utilize graph neural networks for resource allocation by building a graph either from geometric placement of devices \cite{lee2019graph} or from wireless fading \cite{shen2019graph}. Both of these works utilize a simpler GNN architecture than that considered here. In particular, the proposed GNNs in \cite{lee2019graph,shen2019graph} do not use full graph filters as we define in \eqref{eq_graph_conv} and, as a result, do not take higher order neighborhood information into account. Moreover, the previous works tackle simpler resource allocation problems without constraints, as considered by our overall framework.
\end{remark}

\section{Permutation Invariance and Equivariance}\label{sec_equivariance}

Once trained, a GNN can be executed in any network independently of dimension or shape. Indeed, if we are given the filter coefficients to use in \eqref{eqn_gnn_intermediate_features} we can implement the GNN for any graph $\bbH$. This is important for us because the graph $\bbH$ is randomly drawn along with input $\bbx$ from the distribution $m(\bbH,\bbx)$. But it is also important because it allows execution on different networks, i.e., on networks that are drawn from a different distribution $\hhatm(\hbH,\hbx)$. If we draw graphs $\hbH$ and states $\hbx$ from distribution $\hhatm(\hbH,\hbx)$ these can be substituted into \eqref{eq_layer_l_output} to produce hidden layer signals $\hbz_l$ and and outputs that according to the notation in \eqref{eq_gnn_param} we can write as
\begin{align}\label{eq_gnn_param_new_network}
   \bbPhi (\hbH, \hbx; \bbA) = \hbz_L.
\end{align}
Although \eqref{eq_gnn_param_new_network} is just a restatement of \eqref{eq_gnn_param} with different notation, we write it to emphasize that in \eqref{eq_gnn_param_new_network} the graph $\hbH$ and the state $\hbx$ are drawn from a different distribution whereas the filter tensors are the {\it same} in both equations. We say that the tensor that we learn for distribution $m(\bbH,\bbx)$ is transferred to distribution $\hhatm(\hbH,\hbx)$.

Transference of REGNNs can be attempted for any pair of network distributions but we do not expect good performance for transference between arbitrary network pairs. To characterize cases where we do expect good transference performance we will show here that optimal filter tensors are invariant to permutations. We begin by defining permutation matrices of dimension $m$ as those matrices $\bbPi$ that belong to the set
\begin{equation} \label{eqn:permutationSet}
    \varphi = \left\{
        \bbPi \in \{0,1\}^{m \times m}
            \ : \
        \bbPi\, \bbone = \bbone , \
        \bbPi^{T} \bbone = \bbone
    \right\}.
\end{equation}
A permutation matrix $\bbPi$ satisfying the conditions in \eqref{eqn:permutationSet} is one for which the product $\bbPi^{T}\bbv$ reorders the entries of any given vector $\bbv$ and in which the product $\bbPi^{T}\bbM\bbPi$ reorders the rows and columns of any given matrix $\bbM$. We further introduce two assumptions on the permutation invariance of the functions and constraints that define the optimal resource allocation problem in \eqref{eq_problem}.

%
\begin{assumption}\label{assumption_u_zero}
The utility $u_0$ is permutation invariant so that for all permutation matrices $\bbPi\in\varphi$ it holds $u_0(\bbPi^T\bbr)=u_0(\bbr)$ 
\end{assumption}

%
\begin{assumption}\label{assumption_u}
The constraint $\bbu(\bbr)\geq \bb0$ is permutation invariant in the sense that for all $\bbPi\in\varphi$ it holds
\begin{align}
   \bbu\big(\bbr\big)\geq\bb0 \ \iff \ \bbu\big(\bbPi^T\bbr\big)\geq\bb0 .
\end{align} \end{assumption}

%
\begin{assumption}\label{assumption_f}
The reward function $\bbf$ [cf. \eqref{eqn_instantaneous_reward}] is permutation equivariant. I.e., for all permutation matrices $\bbPi\in\ccalP$ it holds 
\begin{align}
   \bbf\Big(\hbp; \, \hbH,\hbx\Big) 
      = \bbPi^T \bbf\Big(\bbp;\, \bbH,\bbx\Big).
\end{align} 
where $\hbH = \bbPi^T \bbH\,\bbPi$ and $\hbx=\bbPi^T\bbx$ are state permutations and and $\hbp=\bbPi^T\bbp$ is a resource allocation permutation.

\end{assumption}

%
Assumption \ref{assumption_u_zero} states that the utility $u_0$ does not change if nodes are reordered. Assumption \ref{assumption_u} states the same for the constraint $\bbu(\bbx)\geq \bb0$. It may be that the components of the vector function $\bbu\big(\bbPi^T\bbx\big)$ are reordered upon permutation, but any constraint that appears in $\bbu(\bbx)$ appears in $\bbu\big(\bbPi^T\bbx\big)$. Assumption \ref{assumption_f} requires that a reordering of the nodes results in a consistent reordering of the reward function. Since the utilities $u_0$ and $\bbu$ are design choices we can enforce Assumptions \ref{assumption_u_zero} and \ref{assumption_u} to hold. Most usual choices for these utilities satisfy these assumptions. Assumption \ref{assumption_f} depends on the physical model of the system. It is not a stringent requirement. All examples in Section \ref{sec_examples} satisfy Assumption \ref{assumption_f}.

We can now state the following theorem.

%
\begin{theorem}\label{theo_invariance}
Consider wireless networks defined by probability distributions $m(\bbH, \bbx)$ and $\hhatm(\hbH, \hbx)$ such that there exists a permutation matrix $\bbPi$ such that if we define $\hbH = \bbPi^T \bbH\,\bbPi$ and $\hbx=\bbPi^T\bbx$ it holds
\begin{align}\label{eqn_theo_invariance_hypothesis}
   \hhatm(\hbH, \hbx) \ =\ \hhatm( \bbPi^T \bbH\,\bbPi, \bbPi^T\bbx)
                      \ =\ m( \bbH,\bbx).
\end{align}
Further assume that Assumptions \ref{assumption_u_zero}, \ref{assumption_u} and \ref{assumption_f} hold. The solutions $\bbA^*$ and $\hbA^*$ of \eqref{eq_gnn_param_problem} for distributions $m(\bbH, \bbx)$ and  $m(\hbH, \hbx)$ are equivalent, 
\begin{align}\label{eqn_theo_invariance}
   \hbA^* \equiv \bbA^*
\end{align} \end{theorem}

%
Theorem \ref{theo_invariance} states permutation {\it invariance} of optimal filter tensors. If two networks are permutations of each other, the respective optimal REGNN filters are the same. Therefore, a REGNN that is trained over the network distribution $m(\bbH, \bbx)$ can be transferred to the network distribution of a permuted network $\hhatm(\hbH, \hbx)$ without loss of optimality. This is a useful property as it is not unreasonable to expect different large scale networks to be close to mutual permutations as we have already mentioned and illustrated in Figure \ref{fig_system}. This implication of Theorem \ref{theo_invariance} is explored numerically in Section \ref{sec_transference}.

Theorem \ref{theo_invariance} is a direct consequence of the fact that, both, REGNNs and Problem \eqref{eq_problem} are {\it equivariant} to permutations as we show in Sections \ref{sec_perm} and  \ref{sec_equivariance_regnns}.

%
\subsection{Permutation equivariance of optimal resource allocation}\label{sec_perm}

A function or policy that demonstrates permutation equivariance is one such that a permutation of inputs results in an equally permuted output. If Assumptions \ref{assumption_u_zero}, \ref{assumption_u} and \ref{assumption_f} hold the following proposition asserts permutation equivariance of resource allocation policies in \eqref{eq_problem}.

%
\begin{proposition}\label{prop_perm} Consider wireless networks defined by probability distributions $m(\bbH, \bbx)$ and $\hhatm(\hbH, \hbx)$ along with resource allocations $\bbp$ and $\hbp$. Assume there exists a permutation matrix $\bbPi$ for which
\eqref{eqn_theo_invariance_hypothesis} holds and that for the same permutation matrix
\begin{align}\label{eq_prop_perm_hypo}
   \hbp(\hbH,\hbx)=\bbPi^T\bbp(\bbH,\bbx).
\end{align}
Define the respective long term rewards $\bbr = \mbE_{m}\big[ \bbf\big(\bbp(\bbH,\bbx); \bbH, \bbx\big) \big]$ and $\hbr = \mbE_{\hhatm}\big[ \bbf\big(\hbp(\hbH,\hbx); \hbH, \hbx\big) \big]$ as per \eqref{eqn_average_reward}. If Assumptions \ref{assumption_u_zero}-\ref{assumption_f} hold, 
\begin{align}\label{eq_prop_perm_result_1}
   u_0(\hbr)=u_0(\bbr),  \quad
   \text{and} \quad
   \bbu\big(\hbr\big)\geq\bb0 \, \iff \, \bbu\big(\bbr\big)\geq\bb0.
\end{align}
In particular, the optimal resource allocations in \eqref{eq_problem} is permutation equivariant in that for any permutation matrix $\bbPi \in \varphi$,
\begin{equation}\label{eq_p_perm}
   \bbp^*\Big(\bbPi^T \bbH\,\bbPi, \bbPi^T \bbx\Big) = \bbPi^T \bbp^*\Big(\bbH,\bbx\Big).
\end{equation} \end{proposition}

%
\begin{myproof} To prove the result in \eqref{eq_prop_perm_result_1} we prove that $\hbr = \bbPi^T\bbr$. This follows readily from their definitions and the hypothesis. Begin by writing 
\begin{align}\label{eq_prop_perm_pf_10}
   \hbr = \int \bbf\Big(\hbp\big(\hbH,\hbx\big); \hbH, \hbx\Big)  d \hhatm(\hbH, \hbx),
\end{align}
and observe that we have assumed $\hbp(\hbH,\hbx)=\bbPi^T\bbp(\bbH,\bbx)$. Substitute this assumption into \eqref{eq_prop_perm_pf_10} to obtain 
\begin{align}\label{eq_prop_perm_pf_20}
   \hbr = \int \bbf\Big(\bbPi^T\bbp\big(\bbH,\bbx\big); \hbH, \hbx\Big)  d \hhatm(\hbH, \hbx),
\end{align}
Implement the change of variables $\hbH \to \bbPi^T \bbH\,\bbPi$ and $\hbx \to \bbPi^T\bbx$. Since the permutation matrices are isometric, this change of variables transforms \eqref{eq_prop_perm_pf_20} into
\begin{align}\label{eq_prop_perm_pf_30}
   \hbr = \int \bbf\Big(\bbPi^T\bbp\big(\bbH,\bbx\big);  \bbPi^T \bbH\,\bbPi, \bbPi^T\bbx\Big)  
   d \hhatm\Big( \bbPi^T \bbH\,\bbPi, \bbPi^T\bbx\Big),
\end{align}
As per assumption \eqref{assumption_f} we know that the function $\bbf$ is permutation equivariant and that, therefore, $\bbf(\bbPi^T\bbp\big(\bbH,\bbx\big);  \bbPi^T \bbH\,\bbPi, \bbPi^T\bbx) =  \bbPi^T\bbf(\bbp\big(\bbH,\bbx\big); \bbH, \bbx)$. As per \eqref{eqn_theo_invariance_hypothesis} we know that the state distributions satify $\hhatm( \bbPi^T \bbH\,\bbPi, \bbPi^T\bbx) = m( \bbH,\bbx)$. Substituting these two facts into \eqref{eq_prop_perm_pf_30} leads to 
\begin{align}\label{eq_prop_perm_pf_40}
   \hbr = \int \bbPi^T\bbf\Big(\bbp\big(\bbH,\bbx\big); \bbH, \bbx\Big)  d m(\bbH, \bbx),
\end{align}
Extracting the permutation matrix from inside the integral and using the definition of $\bbr = \mbE\big[ \bbf\big(\bbp(\bbH,\bbx); \bbH, \bbx\big) \big]$ leads to 
\begin{align}\label{eq_prop_perm_pf_50}
   \hbr = \bbPi^T \bbr.
\end{align}
Given that \eqref{eq_prop_perm_pf_50} holds, the result in \eqref{eq_prop_perm_result_1} follows from direct application of Assumptions \ref{assumption_u_zero} and \ref{assumption_u}. This statement says that permutations of a network and associated permutations of resource allocation functions result in feasible rewards that attain the same utility. Therefore, \eqref{eq_p_perm} holds as a particular case of \eqref{eq_prop_perm_pf_50} for the optimal resource allocation functions $\bbp^*$ and $\hbp^*$. \end{myproof}

%
In Proposition \ref{prop_perm}, we establish that the optimization problem in \eqref{eq_problem} is permutation equivariant and that, in particular, optimal policies are permutation equivariant. We point out that this property for the resource allocation policy follows intuition, as the labeling of the nodes is generally arbitrary--see Remark \ref{remark_w}---and the structure of the policy should indeed reflect that. By identifying such a structural property of the policy we wish to model, we further identify a potentially lower-dimensional class of parameterizations to perform optimization over. A generic parameterization such has the FCNN may contain instances that are permutation equivariant, but it will not hold this property by default. We proceed to establish the permutation equivariance of the class of REGNNS.

%
\subsection{Equivariance of Random Edge Graph Neural Networks}\label{sec_equivariance_regnns}

In comparison to its fully connected counterpart, the REGNN may appear limited by its lack of universal approximation capabilities. However, what we lose in universality we gain in \emph{structure}. That is, in learning the weights of a REGNN in \eqref{eq_gnn_param_problem} we restrict our attention to a class of parameterizations that maintain desirable structure. The convolutional structure of the REGNN architecture allows us to establish the same permutation equivariance property demonstrated for optimal resource allocation policy in Proposition \ref{prop_perm}---a permutation of the underlying graph and input signal of an  REGNN will produce an equally permuted output. This result is stated formally in the following proposition.

%
\begin{proposition}\label{prop_pp_perm} Consider graphs $\bbH$ and $\hbH$ along with signals $\bbx$ and $\hbx$ such that for some permutation matrix $\bbPi$ we have $\hbH = \bbPi^T \bbH\,\bbPi$ and $\hbx = \bbPi^T\bbx$. The output of a REGNN with filter tensor $\bbA$ to the pairs $(\bbH, \bbx)$ and $(\hbH, \hbx)$ are such that
\begin{align}\label{eq_pp_perm_1}
   \bbPhi (\hbH, \hbx; \bbA) = \bbPi^T \bbPhi (\bbH, \bbx; \bbA).
\end{align}
%
\end{proposition}

%
\begin{myproof}  
This is a restatement of \cite[Proposition 2]{gama2019stability}. We sketch a proof assuming the number of filters is $F_l = 1$ at each layer $l$ for completeness. Consider first the layer $l=1$ and take as inputs the permuted node state $\bbz_0 = \hbx = \bbPi^T \bbx$ and permuted graph $\hbH = \bbPi^T \bbH \bbPi$. The output of the first layer for the REGNN is given by \eqref{eq_layer_l_output} as
\begin{align}
\bbz^{\prime}_2 = \bbsigma_1 [  \bbA_1(\hbH) \hbx ]. 
\end{align}
By using the fact that $\bbPi^T \bbPi = \bbPi \bbPi^T = \bbI$ for any permutation matrix $\bbPi$, it follows that $\hbH^k = \bbPi^T \bbH^k \bbPi$. By expanding the term $\bbA_1(\hbH)$ as in \eqref{eq_single_feature_gnn} we obtain
\begin{align}
\bbz^{\prime}_2 &= \bbsigma_1 \left[ \sum_{k=0}^{K_1-1} \alpha_{1,k} \bbPi^T \bbH^k \bbPi \bbPi^T \bbx \right]. \\
&= \bbsigma_1 \left[ \bbPi^T \sum_{k=0}^{K_1-1} \alpha_{1,k}  \bbH^k  \bbx \right] = \bbsigma_1[ \bbA_1(\bbH)\bbx]. 
\end{align}
Because the non-linearity $\bbsigma_1(\cdot)$ is pointwise, it follows that $\bbz^{\prime}_2 = \bbPi^T \bbsigma_1(\bbalpha_1 *_\bbH \bbx) = \bbPi^T \bbz_2$, where $\bbz_2$ is the output of the first layer under unpermuted inputs. As the output of a single layer is permutation equivariant to its input, it follows that the output of the composition of layers $l=1,\dots,L$ is permutation equivarient, i.e. $\bbPhi(\hbH, \hbx; \bbA) = \bbPi^T \bbPhi(\bbH, \bbx; \bbA )$ as stated in \eqref{eq_pp_perm_1}.
%
\end{myproof}

%
Proposition \ref{prop_pp_perm} establishes the permutation equivariance of REGNNs. This structural property comes from the multiplicative relationship between the channel states $\bbH$ and the node states $\bbx$ used in the graph filter in \eqref{eq_graph_conv}. In the context of wireless networks, this implies that a relabelling or reordering of the transmitters in the network will produce an appropriately permutation of the power allocation \emph{without any permutation of the filter weights}. This essential structural property is not satisfied by general FCNNs, in which a restructuring of the network would require an equivalent permutation of the interlayer weights. While their full generality implies that such a permutation equivariant property \emph{can} be satisfied by a FCNN, this property would have to be learned during training. We further note while that alternative parameterizations may also be shown to exhibit a similar permutation equivariance, our numerical study in Section \ref{sec_numerical_results} of REGNNs shows their particular effectiveness in learning for large scale networks.

The results in Propositions \ref{prop_perm} and \ref{prop_pp_perm} lead directly to our primary result in Theorem \ref{theo_invariance} as we formally show next. 

\medskip\begin{myproof}[of Theorem \ref{theo_invariance}] Suppose we are given the tensor $\bbA^*$ that is optimal for \eqref{eq_gnn_param_problem} for distribution $m(\bbH,\bbx)$. This tensor produces a resource allocation $\bbPhi (\bbH, \bbx; \bbA^*)$ and an optimal reward $\bbr^*$. Permute the GNN inputs with $\bbPi$ to produce the resource allocation $\bbPhi (\hbH, \hbx; \bbA^*)$, which according to the result \eqref{eq_pp_perm_1} of Proposition \ref{prop_pp_perm} satisfies
\begin{align}\label{eq_theo_perm_pf_10}
   \bbPhi (\hbH, \hbx; \bbA^*) = \bbPi^T \bbPhi (\bbH, \bbx; \bbA^*).
\end{align}
From this fact and the hypotheses of Theorem \ref{theo_invariance}, which we are trying to prove, we have that the hypotheses of Proposition \ref{prop_perm} hold. Thus, the reward
\begin{align}\label{eq_theo_perm_pf_20}
   \hbr = \mbE_{\hhatm}\Big[ \bbf\Big(\bbPhi (\hbH, \hbx; \bbA^*); \hbH, \hbx\Big) \Big]
\end{align}
is feasible in problem \eqref{eq_gnn_param_problem} with state distribution $\hhatm(\hbH,\hbx)$ and attains utility $u(\hbr)=u(\bbr^*)$. 

Consider now the optimal tensor $\hbA^*$ that is optimal for \eqref{eq_gnn_param_problem} for distribution $\hhatm(\hbH,\hbx)$. This tensor produces a resource allocation $\bbPhi (\hbH, \hbx; \hbA^*)$ and an optimal reward $\hbr^*$. Repeating the argument that leads to \eqref{eq_theo_perm_pf_10} and \eqref{eq_theo_perm_pf_20} we conclude that the reward
\begin{align}\label{eq_theo_perm_pf_30}
   \bbr = \mbE_{\hhatm}\Big[ \bbf\Big(\bbPhi (\bbH, \bbx; \hbA^*); \bbH, \bbx\Big) \Big]
\end{align}
is feasible in problem \eqref{eq_gnn_param_problem} with state distribution $m(\bbH,\bbx)$ and attains utility $u(\bbr)=u(\hbr^*)$. 

Since we know that $\hbr^*$ is optimal for distribution $\hhatm(\hbH,\hbx)$ and that $\hbr$ is simply feasible we must have
\begin{align}\label{eq_theo_perm_pf_40}
   u(\hbr^*) \geq u(\hbr) = u(\bbr^*),
\end{align}
where the second equality follows from Proposition \ref{prop_perm} as already shown. Likewise, since $\bbr^*$ is optimal for distribution $m(\bbH,\bbx)$ and $\bbr$ is feasible, it must be that
\begin{align}\label{eq_theo_perm_pf_50}
   u(\bbr^*) \geq u(\bbr) = u(\hbr^*).
\end{align}
For \eqref{eq_theo_perm_pf_40} and \eqref{eq_theo_perm_pf_50} to hold the inequalities must be equalities. Thus, $\bbA^*$ must be optimal for distribution $\hhatm(\hbH,\hbx)$ and, conversely, $\hbA^*$ must be optimal for distribution $m(\bbH,\bbx)$. The result in \eqref{eqn_theo_invariance} is implied. \end{myproof}

%
\begin{remark}[Permutation Matrix is Unknown]\label{remark_w}\normalfont The permutation equivariance results in Propositions \ref{prop_perm} and \ref{prop_pp_perm}, which imply the permutation invariance result of Theorem \ref{theo_invariance}, do not require knowledge of the permutation $\bbPi$ that equalizes the distributions $m(\bbH, \bbx)$ and $\hhatm(\hbH, \hbx)$ in \eqref {eqn_theo_invariance_hypothesis}. This is worth remarking because if the permutation is known, designing operators that are permutation invariant is elementary -- just undo the permutation and apply the corresponding operator. REGNNs achieve permutation invariance without knowing the permutation that relates $m(\bbH, \bbx)$ and $\hhatm(\hbH, \hbx)$. This is as it should be to enable transference---see Section \ref{sec_transference}. \end{remark}

%
\begin{remark}[Homogeneity vs Heterogeneity]\label{remark_w}\normalfont However advantageous, permutation invariance hinders the ability to handle heterogeneous agents. Consider, for example, a weighted utility $u_0(\bbr) = \bbw^T\bbr$ that assigns different weights to the rewards of different agents but assume that otherwise, the hypotheses of Propositions \ref{prop_perm} and \ref{prop_pp_perm} and Theorem \ref{theo_invariance} hold. This utility violates Assumption \ref{assumption_u_zero} but this assumption is not needed in Proposition \ref{prop_pp_perm}. Thus, we know that outputs of the REGNN are permutation equivariant [cf. \eqref{eq_pp_perm_1}] and that, in particular, the resource allocations satisfy $\hbr = \bbPi^T\bbr$ [cf. \eqref{eq_prop_perm_pf_50}]. We then have that $u_0(\hbr) = u_0(\bbPi^T\bbr) = \bbw^T\bbPi^T\bbr$ which equivocates the weight assignment.  \end{remark}

%

%
\section{Primal-Dual Learning}\label{sec_primal_dual}

To find the optimal filter tensor $\bbA^*$ we find a saddle point of the Lagrangian associated with the optimization problem in \eqref{eq_gnn_param_problem}. To define the Lagrangian let $\bblam$ be a multiplier associated with the constraint $\bbr = \mbE[\bbf(\bbPhi(\bbH,\bbx;\bbA);\bbH,\bbx)]$ and $\bbmu\geq\bb0$ a multiplier associated with the constraint $\bbu(\bbr)\geq\bb0$. Recall that the constraint $\bbPhi(\bbH,\bbx; \bbA) \in \ccalP (\bbH,\bbx)$ in \eqref{eq_gnn_param_problem} is addressed in the output layer of the REGNN and is thus not considered in the Lagrangian function. The Lagrangian of \eqref{eq_gnn_param_problem} is the following weighted combination of objective and constraints
\begin{align} \label{eqn_lagrangian}
   &\ccalL(\bbA,\bbr,\bblam,\bbmu) =            \\ \nonumber &\qquad
             u_0(\bbr) 
           + \bblam^T \bigg[\mbE\Big[\bbf\Big(\bbPhi\big(\bbH,\bbx;\bbA\big);\bbH,\bbx\Big)\Big] - \bbr\bigg]
           + \bbmu^T\bbu(\bbr) .  
\end{align}
A saddle point of the Lagrangian in \eqref{eqn_lagrangian} is a primal dual pair $(\bbA,\bbr)^\dagger, (\bblam,\bbmu)^\dagger$ such that for all variables $(\bbA,\bbr)$, and $(\bblam,\bbmu)$ in a sufficiently small neighborhood we have
\begin{align} \label{eqn_saddle_point}
   \ccalL\Big(\bbA,\bbr,(\bblam,\bbmu)^\dagger\Big) 
      \leq \ccalL\Big((\bbA,\bbr)^\dagger,(\bblam,\bbmu)^\dagger\Big) 
         \leq \ccalL\Big((\bbA,\bbr)^\dagger,\bblam,\bbmu\Big) .
\end{align}
Namely, the saddle point is locally maximal on the primal variables and locally minimal in the dual variables. There several saddle points satisfying \eqref{eqn_saddle_point} because the Lagrangian $\ccalL(\bbA,\bbr,\bblam,\bbmu)$ in \eqref{eqn_lagrangian} is associated to a nonconvex optimization problem \cite[Ch. 5]{bertsekas1997nonlinear}. We know that one of them contains the optimal tensor $\bbA^*$. Since a global search is intractable we are going to settle for a local search. We remark that this search need not find the optimal tensor $\bbA^*$ but we have observed good empirical results. 

The primal-dual method we use here alternates between gradient descent steps in the dual variables $(\bblam,\bbu)$ with gradient ascent steps in the primal variables $(\bbA, \bbx)$. In specific, let $k$ be an iteration index and $\eps$ denote a stepsize. The primal update on the reward function $\bbr$ is $\bbr_{k+1} = \bbr_k + \eps\nabla_\bbr\ccalL(\bbA_k,\bbr_k,\bblam_k,\bbmu_k)$, which, using the explicit Lagrangian expression in \eqref{eqn_lagrangian} results in the update
\begin{align} \label{eqn_reward_update}
   \bbr_{k+1} = \bbr_k 
        + \eps\,\Big[\, \nabla_{\bbr}\,u_0(\bbr_k) 
                            + \big[\nabla_{\bbr}\,\bbu(\bbr_k)\big] \bbmu_k 
                                   - \bblam_k \,\Big] .
\end{align}
Similarly, the update in the dual variable $\bbmu$ takes the form $\bbmu_{k+1} = [\bbmu_k - \eps\nabla_{\bbmu}\ccalL(\bbA_k,\bbr_k,\bblam_k,\bbmu_k)]^+$ where we add a projection on the nonnegative orthant to account for the fact that $\bbmu$ is nonnegative. Since the Lagrangian is linear in the dual variables the gradient is simple to compute and the update reduces to
\begin{align} \label{eqn_mu_update}
   \bbmu_{k+1} = \Big[\, \bbmu_k  + \eps\, \bbu(\bbr_k)  \, \Big]^+.
\end{align}
The updates in \eqref{eqn_reward_update} and \eqref{eqn_mu_update} are both relatively easy to carry out as they depend on the utility funtions $u_0$ and $\bbu$. Taking gradients with respect to the filter tensor $\bbA$ and the multiplier $\bbLam$ is more challenging because of the expectation operators. The gradient descent update with respect to $\bblam$, for example, takes the form
\begin{align} \label{eqn_lambda_update_with_expectation}
  \bblam_{k+1} = \bblam_k - \eps\, \mbE \Big[\,
                    \bbf\Big(\bbPhi\big(\bbH,\bbx;\bbA_k\big);\bbH,\bbx\Big)\, - \bbr_{k}\Big] .
\end{align}
This is an update that can't be computed if the distribution $m(\bbH,\bbx)$ is unknown. This is a problem that is resolved by using stochastic updates in which we sample a realization $(\bbH_k,\bbx_k)$ and update $\bblam_k$ according to
\begin{align} \label{eqn_lambda_update}
  \bblam_{k+1} = \bblam_k - \eps\, \Big[\,
                    \bbf\Big(\bbPhi\big(\bbH_k,\bbx_k;\bbA_k\big);\bbH_k,\bbx_k\Big)\, - \bbr_k \Big] .
\end{align}
The same idea of stochastic updates is used when taking gradients with respect to the filter tensor. This yields the filter update
\begin{align} \label{eqn_filter_update}
  \bbA_{k+1} = \bbA_k + \eps \Big[\nabla_{\bbA}\,
                   \bbf\Big(\bbPhi\big(\bbH_k,\bbx_k;\bbA_k\big);\bbH_k,\bbx_k\Big)\Big] \bblam_k
\end{align}
It is germane to point out that to implement \eqref{eqn_lambda_update} we do not need to know the model $\bbf$ mapping resource allocations and network states to user rewards. It suffices to observe the state of the network $(\bbH_k,\bbx_k)$ and implement the resource allocation $\bbPhi(\bbH_k,\bbx_k;\bbA_k)$ according to the current filter tensor iterate $\bbA_k$. We can then observe the reward outcome $\bbf(\bbPhi(\bbH_k,\bbx_k;\bbA_k);\bbH_k,\bbx_k)$ and use it to implement the Lagrange multiplier update in \eqref{eqn_lambda_update}.

The same is not quite true for the update in \eqref{eqn_filter_update} because it requires gradients of $\bbf$ which cannot be directly queried by probing the system. This is a challenge that also arises in policy gradient methods where it is resolved with the introduction of randomized policies \cite{sutton2000policy}. Mimicking that approach we reinterpret $\bbPhi(\bbH,\bbx;\bbA)$ as the parameter of probability distribution $\bbPsi(\bbH,\bbx;\bbA)$. A likelihood ratio identity allows us to express the gradient in \eqref{eqn_filter_update} as
\begin{align}\label{eq_policy_gradient}
\nabla_{\bbA} \mathbb{E} &\Big[ \bbf\Big(\bbPhi\big(\bbH,\bbx;\bbA_k\big);\bbH,\bbx\Big) \Big] =  \\
&\E\Big[ \bbf\Big(\bbPhi\big(\bbH,\bbx;\bbA_k\big);\bbH,\bbx\Big) \nabla_{\bbA} \log \bbPsi(\bbH,\bbx;\bbA_k)^T\Big]. \nonumber
\end{align}
The substitution in \eqref{eq_policy_gradient} is useful in that it replaces the gradient of the expectation of $\bbf$ with the expectation of the gradient of $\log\bbPsi$. As in the stochastic update in \eqref{eqn_lambda_update}, we can estimate the gradient with sampled realization $(\bbH_k, \bbx_k)$, while the gradient of $\log\bbPsi$ can be calculated assuming a given distribution $\bbPsi(\bbH_k, \bbx_k; \bbA_k)$. This results in the stochastic policy for the filter tensor $\bbA$
\begin{align} \label{eqn_filter_update_s}
  &\bbA_{k+1} = \bbA_k + \\ & \eps 
                 \Big[  \bbf\Big(\bbPhi\big(\bbH_k,\bbx_k;\bbA_k\big);\bbH_k,\bbx_k\Big) \nabla_{\bbA} \log \bbPsi(\bbH_k,\bbx_k;\bbA_k)^T \bblam_k \Big]. \nonumber
\end{align}
We stress that \eqref{eqn_filter_update_s} is model-free because, like \eqref{eqn_lambda_update}, it is computed only by probing $\bbf\Big(\bbPhi\big(\bbH_k,\bbx_k;\bbA_k\big);\bbH_k,\bbx_k\Big)$ and thus does not need explicit knowledge of $\bbf$ or the state distribution $m(\bbH,\bbx)$.

We train REGNNs using the updates in \eqref{eqn_reward_update}, \eqref{eqn_mu_update}, \eqref{eqn_lambda_update}, and \eqref{eqn_filter_update_s}. The resulting scheme is summarized in Algorithm \ref{alg:learning}. To initialize the training process we specify the REGNN architecture. We use the shorthand $\text{REGNN}[L, (F_1,K_l),\ldots,(F_L, K_L)]$ to signify a REGNN with $L$ layers in which the $l$th layer contains $F_l$ features generated by filters of length $K_l$.  After initializing REGNN filter tensor $\bbA_0$ and other primal and dual variables in Step 2, we perform our learning iterations in Step 3. Each learning iteration $k$ consists of first computing of drawing samples of states $\bbH$ and $\bbx$ and probing the system $\bbf\Big(\bbPhi\big(\bbH_k,\bbx_k;\bbA_k\big);\bbH_k,\bbx_k\Big)$ using the current REGNN policy specified by $\bbA_k$ in Step 4. In Step 5 we perform the primal-dual gradient updates in \eqref{eqn_reward_update}, \eqref{eqn_mu_update}, \eqref{eqn_lambda_update}, and \eqref{eqn_filter_update_s} using the probed performance. The process is repeated until convergence.

%
{\linespread{1.0}
\begin{algorithm}[t] \begin{algorithmic}[1]
\STATE \textbf{Parameters:} $\text{REGNN}[L, (F_1,K_l),\ldots,(F_L, K_L)]$
\STATE \textbf{Input:} Initial states $\bbA_0, \bbr_0, \bblambda_0, \bbmu_0$
\FOR [main loop]{$k = 0,1,2,\hdots$}
     \STATE Sample states $(\bbH_k, \bbx_k) \sim m(\bbH,\bbx)$. \\ Probe $\bbf\big(\bbPhi\big(\bbH_k,\bbx_k;\bbA_k\big);\bbH_k,\bbx_k\big)$.
      \STATE Update primal and dual variables [\eqref{eqn_reward_update}, \eqref{eqn_mu_update}, \eqref{eqn_lambda_update}, \eqref{eqn_filter_update_s}]
      \footnotesize{
         \begin{align}
	   &\bbr_k 
        + \eps\,\Big[\, \nabla_{\bbr}\,u_0(\bbr_k) 
                            + \big[\nabla_{\bbr}\,\bbu(\bbr_k)\big] \bbmu_k 
                                   - \bblam_k \,\Big]  \nonumber \\                    
	&\Big[ \bbmu_k  + \eps \bbu(\bbr_k)   \Big]^+\nonumber \\	
	&\bblam_k - \eps \Big[\,
                    \bbf\Big(\bbPhi\big(\bbH_k,\bbx_k;\bbA_k\big);\bbH_k,\bbx_k\Big)\, - \bbr_k \Big]	,\nonumber \\
		 &\bbA_k + \eps 
               \Big[  \bbf\Big(\bbPhi\big(\bbH_k,\bbx_k;\bbA_k\big);\bbH_k,\bbx_k\Big) \nabla_{\bbA} \log \bbPsi(\bbH_k,\bbx_k;\bbA_k)^T \bblam_k \Big] \nonumber
\end{align}}
\ENDFOR
\end{algorithmic}
\caption{Primal-Dual REGNN Training }\label{alg:learning} \end{algorithm}

%


\begin{figure}
\centering
\includegraphics[height=.22\textheight, width = \linewidth]{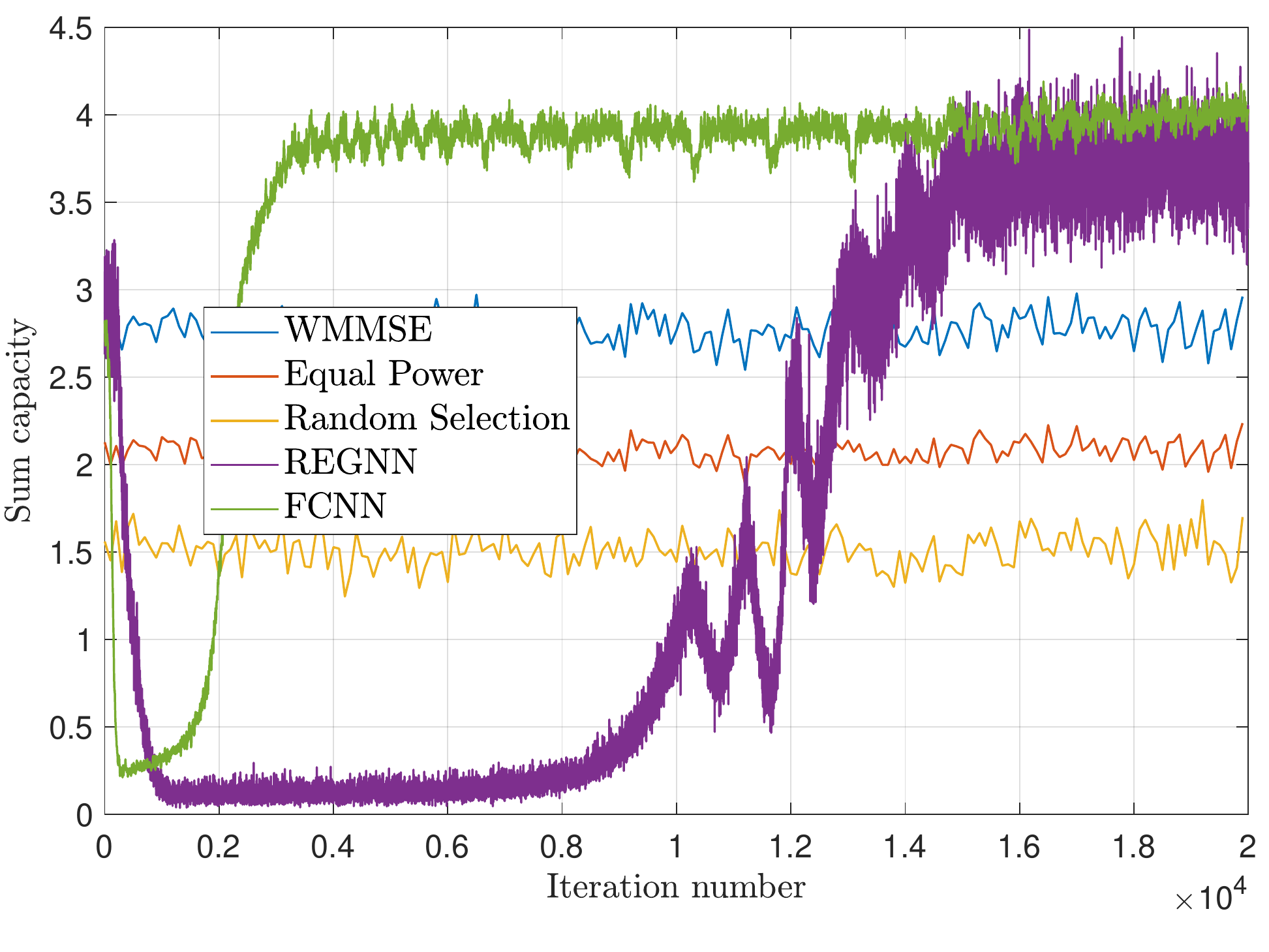}
\caption{Performance comparison during training of REGNN for $m=20$ pairs. With only $q=40$ parameters, the REGNN outperforms the WMMSE algorithm and matches the performance of the FCNN.}
\label{fig1}
\end{figure}

%
\section{Numerical Results} \label{sec_numerical_results}

In this section, we provide a numerical study of the performance of resource allocation policies that parameterized with REGNNs and trained with the model-free primal-dual learning method. We simulate the performance of the policy on a number of canonical resource allocation functions that take the form of \eqref{eq_problem} and compare against existing heuristic approaches. Where applicable, we point out the compared heuristics that rely on accurate model knowledge to be implemented, which as discussed in Section \ref{sec_primal_dual}, is \emph{not} required to implement the primal-dual learning method. We study the canonical problem of binary power control between $m$ transmitter/receiver pairs over an AWGN channel with noise power $\sigma^2$ and interference---see the first example in Section \ref{sec_examples} for a discussion of this problem. In addition to maximizing the sum-rate capacity, a practical constraint of interest is a maximum average power budget $P_{\max}$ to be shared between transmitters connected to a common power supply. The complete resource allocation problem can be written as 
\begin{align} \label{eq_power_control}
   P^* & :=   \max_{\bbp(\bbH,\bbx),\bbr} \  \sum_{i=1}^m r_i,             \\
        &  \st            \         r_i      =  \E \Big[  \log \bigg(1 + \frac{|h_{ii}|^2 p_i(\bbH)}
              {\sigma^2 + \sum_{j \neq i}| h_{ji}|^2 p_j(\bbH)}\bigg) \Big], \nonumber  \\
        &    \qquad                      \E \left[\mathbf{1}^T \bbp(\bbH,\bbx)\right] \leq P_{\max}, \quad 
                                        \bbp(\bbH,\bbx) \in  \{0, p_0\}^m.   \nonumber%
\end{align}
As discussed in Section \ref{sec_examples}, the problem in \eqref{eq_power_control} does not utilize any node state $\bbx$ and associated cost constraint. This is however an instructive problem to study, as it is well studied and has numerous developed heuristic solutions with which the compare as baselines. Observe also that the power allocation is a binary selection of transmitting with power $p_0$ or not transmitting.

\subsection{Ad-hoc networks}\label{sec_adhoc}

We begin studying the performance of the model-free training of an REGNN in  the wireless ad-hoc, or paired network. For all sets of simulations, we construct the ad-hoc wireless network as follows. For a set of $m$ pairs, we construct a random geometric graph by dropping transmitter $i$ uniformly at random at location $\bbt_i \in [-m,m]^2$, with its paired receiver at location $\bbr_i  \in [\bbt_i - m/4, \bbt_i + m/4]^2$ around its paired transmitter---see, e.g., Figure \ref{fig_systema} for an example. Given the geometric placements, the complete fading channel state between transmitter $i$ and receiver $j$ is composed of $h_{ij} = h^{p}_{ij} h^{f}_{ij}$, where $h^{p}_{ij}$ is a constant path-loss gain and $h^f_{ij}$ is the time varying fast fading. The path loss is related to the geometric distance as $h^p_{ij} = \|\bbt_i - \bbr_j\|^{-2.2}$ and the fast fading $h^f_{ij}$ is drawn randomly from a standard Rayleigh distribution at each scheduling cycle and the noise power is initially fixed at $\sigma^2 = 1$.

In employing the primal dual learning method in Algorithm \ref{alg:learning}, we consider the model free version in which gradients are estimated via the policy gradient approximation. We construct an REGNN architecture with $L=8$ hidden layers, each with $F_l=1$ graph filters of length $K_l = 5$ and a standard ReLu non-linear activation function i.e. $\bbsigma(\bbz) = [\bbz]_+$. The final layer is passed through a sigmoid function to normalize the outputs, which are then used as the parameter of a the policy distribution  $\bbPsi(\bbH,\bbx;\bbA)$---chosen here as a Bernoulli distribution.  The primal dual method is performed with a geometrically decaying step size for dual updates and the ADAM optimizer \cite{kingma2014adam} for the primal updates.

In general, we make our comparisons against existing heuristic methods for solving \eqref{eq_power_control}. We primarily consider (i) the popular WMMSE heuristic \cite{shi2011iteratively} as a baseline, while also making comparisons against naive heuristics that either (ii) assign equal power $P_{\max}/m$ to all users or (iii) randomly select $P_{\max}/p_{0}$ users to transmit with full power. Furthermore, we simulate the learning and performance of the convolutional REGNN architecture to a fully connected neural network (FCNN) for medium scale networks. In Fig. \ref{fig1}, we show the performance, or sum-capacity, achieved throughout the learning process of the REGNN and FCNN trained with the primal-dual learning method and the performance of the three heuristic baselines for a medium scale wireless system with $m=20$ pairs. It can be observed that both the REGNN and FCNN outperform the performance of WMMSE for the medium scale system. We stress that this performance is obtained by the NNs using the model-free gradients, meaning that knowledge of the capacity function was not assumed. Explicit knowledge of capacity functions is needed, however, for the WMMSE algorithm. 

Observe that the REGNN, with only $q=40$ parameters, matched the performance of the FCNN with two fully connected layers of size 64 and 32 for a total of $q=20\times64 + 64\times32 + 32\times20\approx4000$ parameters---a 100 factor increase than that used by the REGNN. We also point out that, while $20000$ iterations are needed for convergence here, and it can be further observed in Fig. \ref{fig1} that the REGNN requires more iterations than the FCNN to achieve stronger performance. We first stress that the issue of number of iterations until convergence is not critical in this setting, as the training is done \emph{offline}; the runtime complexity of the REGNN is lower than that of a FCNN. The proceeding simulation experiments demonstrate that the REGNN is capable of scaling to larger systems. In Section \ref{sec_transference}, we will demonstrate how the REGNN is moreover able transfer to networks of different configurations and sizes. This could not be possible with a FCNN, whose input dimension is of a fixed size. Therefore, while in the particular case shown in Fig. \ref{fig1}, it takes longer to train the REGNN, it can be considered overall much slower to train a FCNN, as it necessarily needs to be retrained for each new configuration of the network.

\begin{figure}
\centering
\includegraphics[height=.22\textheight, width = \linewidth]{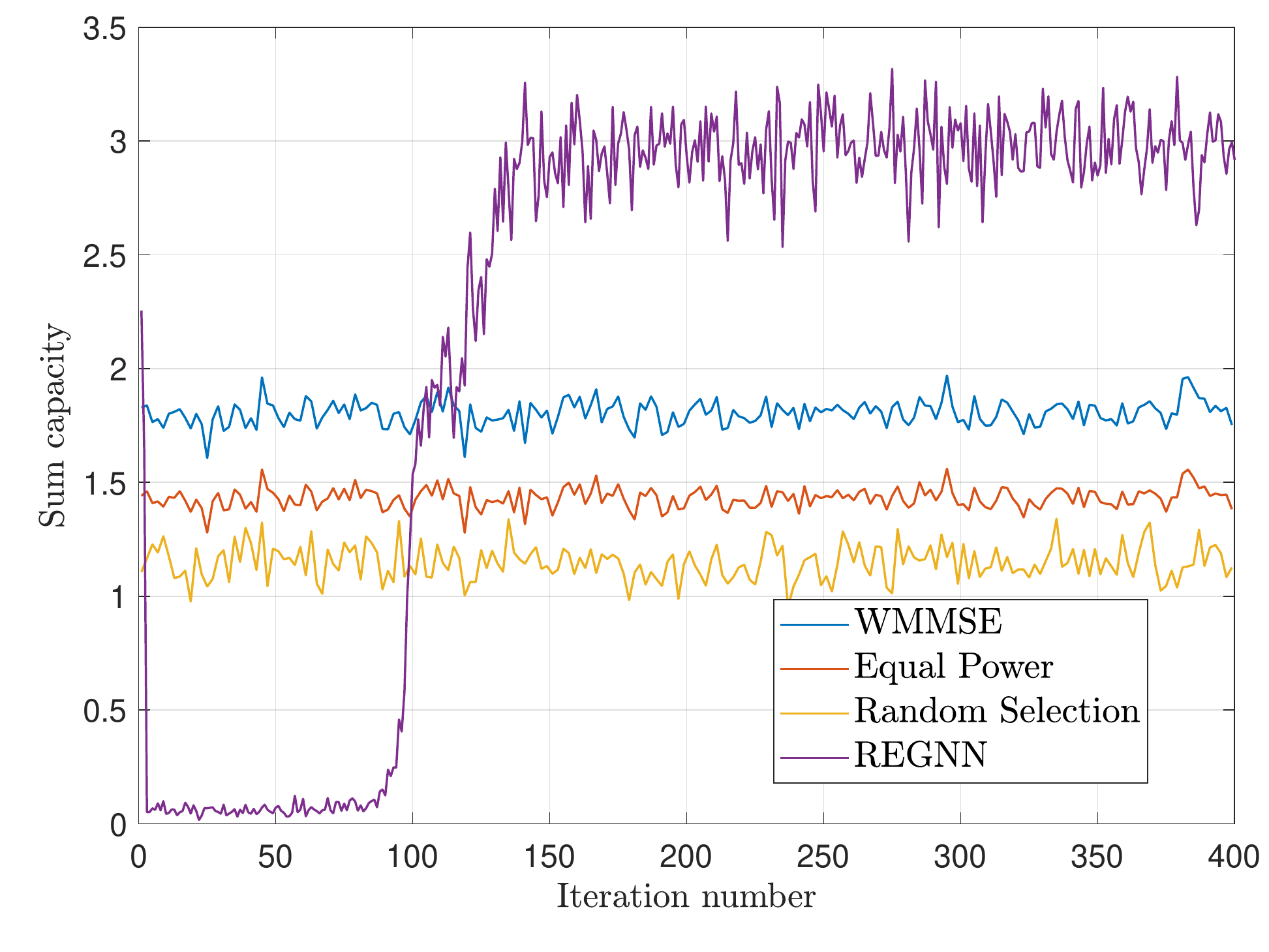}
\caption{Performance comparison during training of REGNN for $m=50$ pairs. With only $q=40$ parameters, the REGNN strongly outperforms the WMMSE algorithm.}
\label{fig2}
\end{figure}

\begin{figure}
\centering
\includegraphics[height=.22\textheight, width = \linewidth]{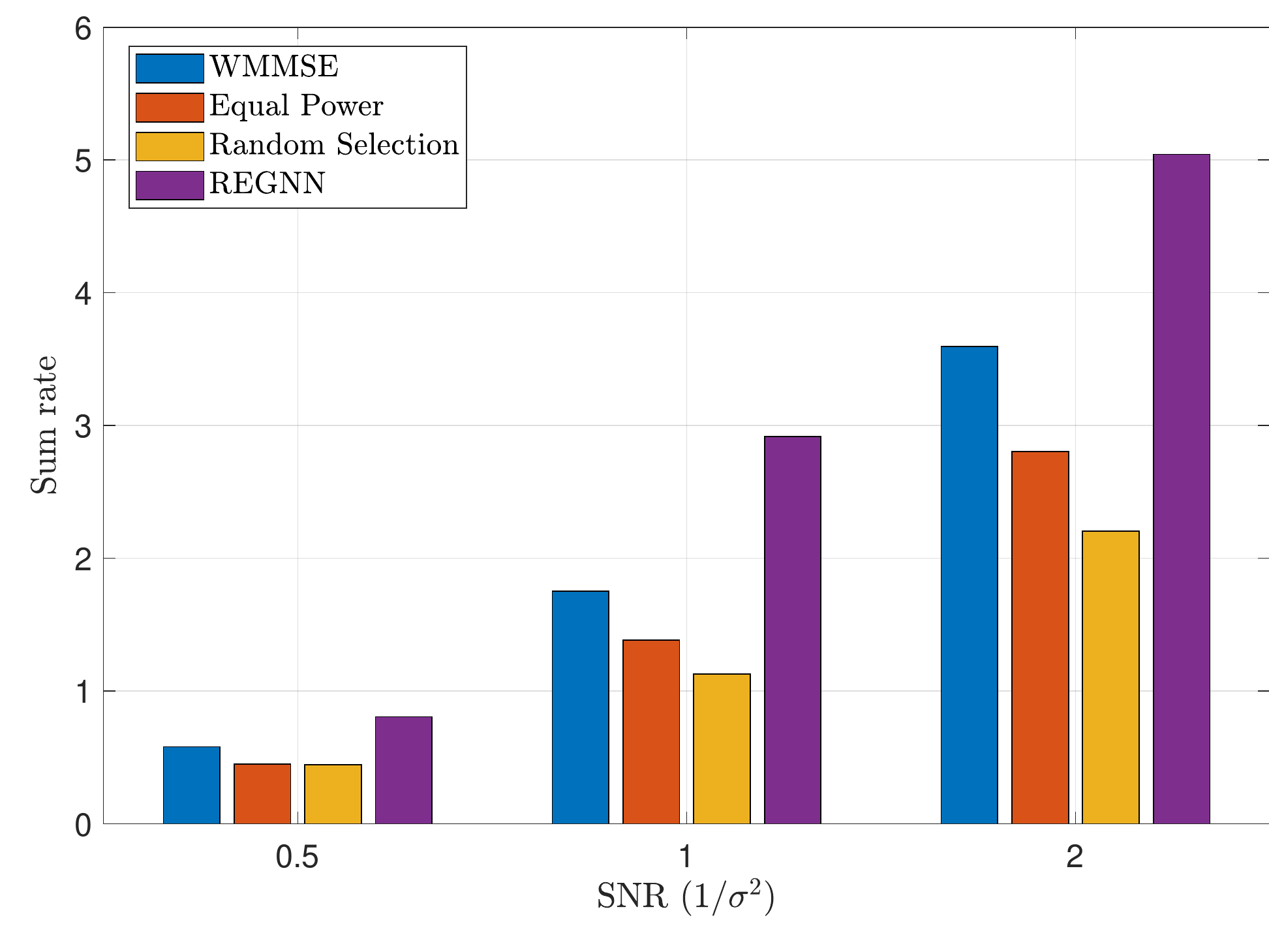}
\caption{Performance comparison of an REGNN against baselines in different SNR regimes. The average SNR is controlled by the AWGN noise parameter $1/\sigma^2$, larger values reflect larger average SNR. The REGNN has bigger gain over WMMSE in higher SNR regimes.}
\label{snr_compare}
\end{figure}

In Figure \ref{fig2}, we show the performance while learning a REGNN in a larger scale system with $m=50$ transmitter/receiver pairs. At this scale, the parameter dimension of the FCNN makes it challenging to train; the input dimension of channel states is $2500$. Here, we see that the learned REGNN substantially outperforms all three heuristics, including WMMSE. Observe that while WMMSE achieves a sum-capacity of roughly 2.7 in the medium scale system, the algorithm performs worse when more transmitters are added, obtaining a sum-rate of only 1.8. The REGNN, meanwhile, is able to still achieve a sum-rate of 3.0, all while only learning $40$ parameters. In Fig. \ref{snr_compare}, we evaluate the gains of the learned REGNN over the baselines in different SNR regimes. To vary the average SNR, we set the AWGN noise parameter $1/\sigma^2$  to vary between $\{0.5, 1, 2 \}$, where a larger value reflects a higher average SNR. As can be seen, the REGNN has substantial gain over all methods in the middle and larger SNR regimes, but only matches the performance of WMMSE when the average SNR value is small. This follows from the fact that WMMSE is known to approach the optimal value of the sum-rate maximization problem as the SNR decreases \cite{shi2011iteratively} and corroborates similar findings in comparisons with other machine learning-based methods, e.g. \cite{liang2018towards}. 

\begin{figure}
\centering
\includegraphics[height=.22\textheight, width = \linewidth]{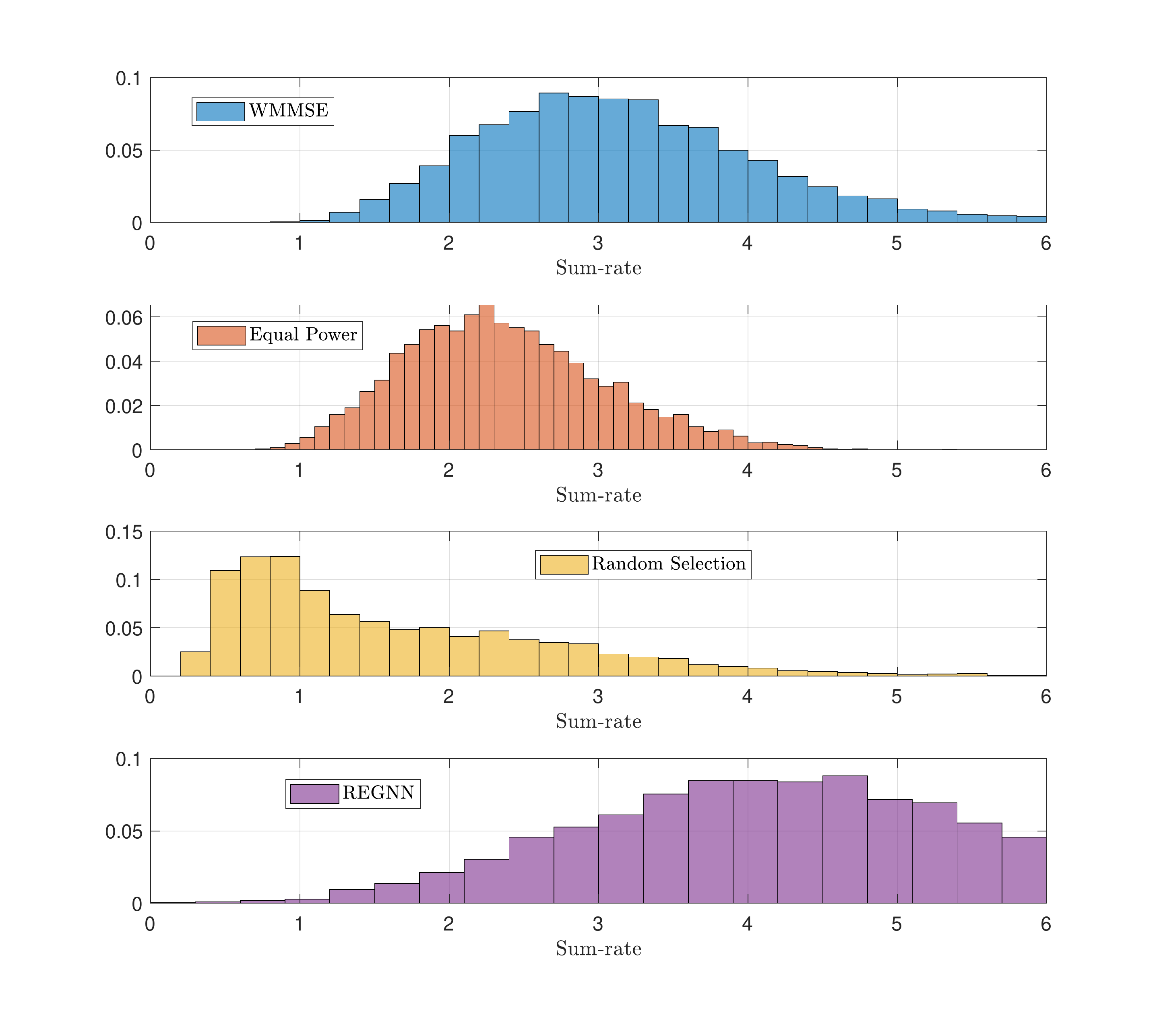}
\caption{Performance comparison of an REGNN that was trained in Figure \ref{fig2} in another randomly drawn network of equal size. Presented is the empirical distribution of the sum-rate achieved over many random iterations for all heuristic methods.}
\label{fig3}
\end{figure}

\subsection{Transference}\label{sec_transference}

As previously discussed, we are interested in exploring the generalization abilities of an REGNN learned over some fixed network. Recall that the filter-bank structure of an REGNN in \eqref{eq_gnn_param} allows the same neural network to receive inputs of varying input dimension, or network size. The permutation invariance result in Theorem \ref{theo_invariance} demonstrates that an REGNN trained on one network performs as well on permutation of that network. As we don't expect to see exact permutations in practice, here we numerically investigate the transference capabilities on randomly drawn networks of fixed density.  Consider the REGNN leaned in the previous experiment in Figure \ref{fig2}. As an instructive example, consider another randomly drawn network of 50 pairs as shown in Figure \ref{fig3}. The performance of the REGNN trained in Figure \ref{fig2} over many random iterations is shown here compared to the heuristics as an empirical histogram of sum-rates over all random iterations. We see that the same parameterization learned for one network performs well with another network. Intuitively speaking, this relates to the stability and permutation equivariance of GNNs because random networks of size 50 may be close to each other in expectation.

Another comparison of interest here is the relative performance of a REGNN trained on a network of size $m=50$ with an REGNN trained on a network of $m^{\prime} > m$. In Figure \ref{fig_75_test} we show a histogram of the sum-rate performance over 50 randomly generated networks of size $m^\prime = 75$ and $m^{\prime} = 100$. For a set of $m^{\prime}$ pairs, we construct a random geometric graph by dropping transmitter $i$ uniformly at random at location $\bbt_i \in [-m \sqrt{m^{\prime}/m},m \sqrt{m^{\prime}/m}]^2$, with its paired receiver at location $\bbr_i  \in [\bbt_i - m/4, \bbt_i + m/4]^2$ around its paired transmitter. This is done to keep the density of the network constant as the number of transceiver pairs grows.

The performance for each random network is itself evaluated over 100 separate fast fading samples. As can be seen, the performance of the REGNN trained on the smaller network of size $m=50$ almost matches the performance of an REGNN trained on a network of size 75. The same procedure is performed for networks of size $m^{\prime}=100$. In Figure \ref{fig_100_test}, we show the performance of the REGNN trained on a network of size 50 against the performance of a network trained on a network of size 100 on random networks of size 100. Again, the performance of the REGNN trained on the smaller network only slightly degrades relative to the REGNN trained on the larger network. This highlights a potential to train REGNNs on smaller networks to later be implemented on larger networks. We point out that this is a powerful property for practical learning for such systems, as we can potentially train our neural networks on smaller systems when larger networks are either unavailable during training or when computational expense is prohibitive. 

\begin{figure}[t]
\centering
\includegraphics[height=.22\textheight, width = \linewidth]{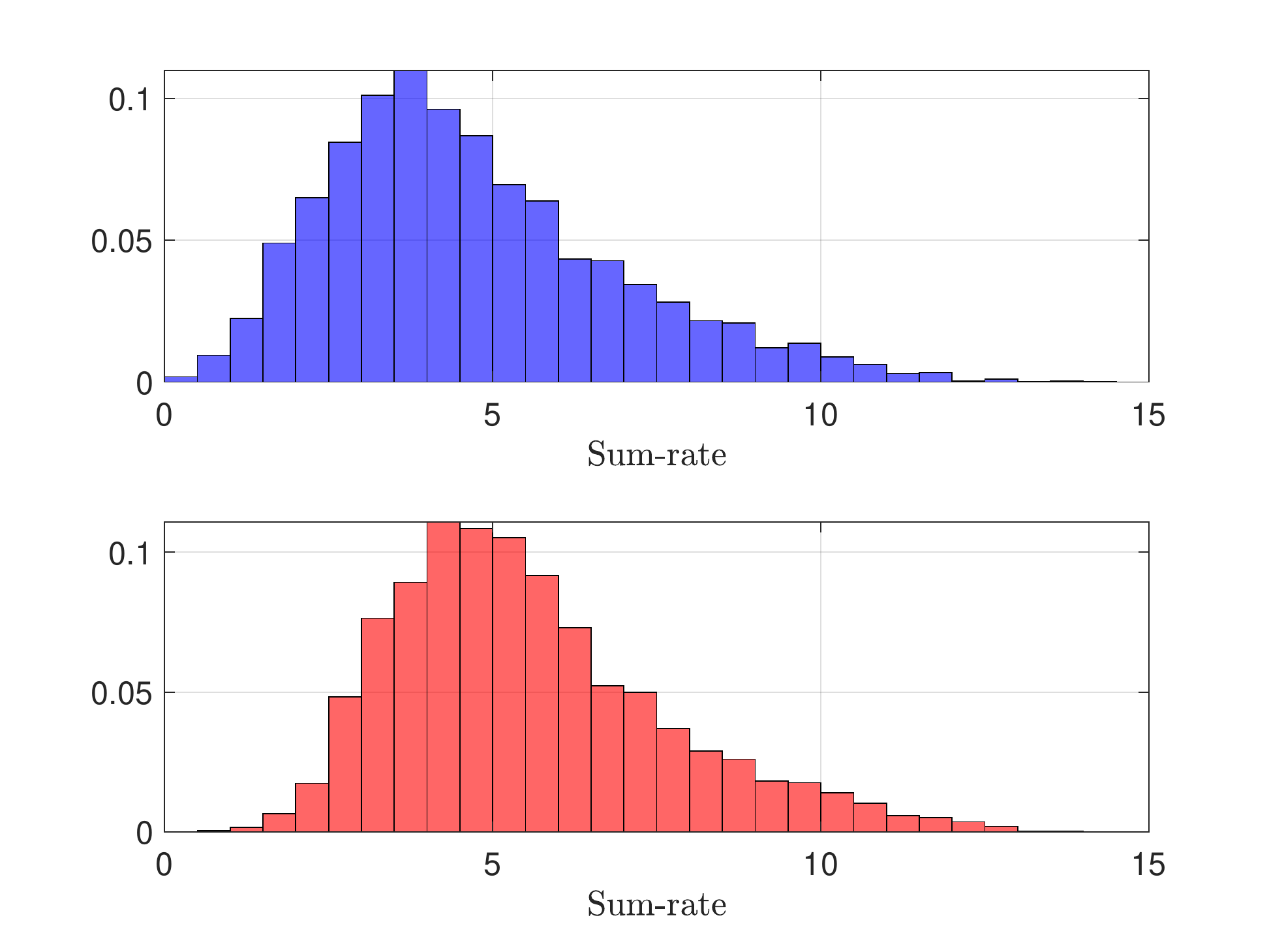}
\caption{Empirical histogram of sum rates obtained by (top, blue) REGNN trained on network of size $m=50$ and (bottom, red) REGNN trained on network of size $m^{\prime}=75$ on 50 randomly drawn networks of size of $m^{\prime}=75$. The REGNN trained on the smaller network closely matches the performance of the REGNN trained on the larger network}
\label{fig_75_test}
\end{figure}

\begin{figure}[t]
\centering
\includegraphics[height=.22\textheight, width = \linewidth]{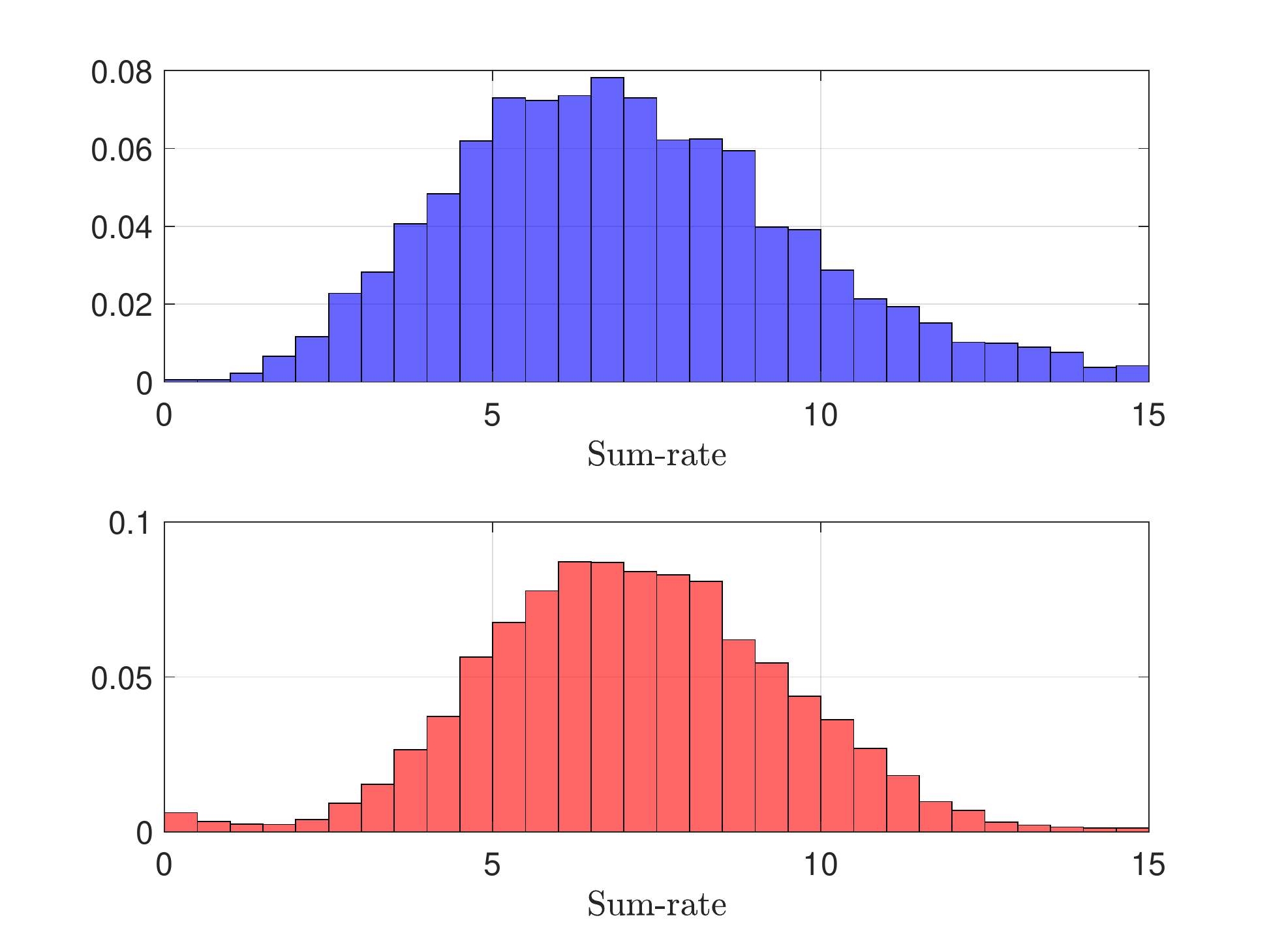}
\caption{Empirical histogram of sum rates obtained by (top, blue) REGNN trained on network of size $m=50$ and (bottom, red) REGNN trained on network of size $m^{\prime}=100$ on 50 randomly drawn networks of size of $m^{\prime}=100$. The REGNN trained on the smaller network closely matches the performance of the REGNN trained on the larger network}
\label{fig_100_test}
\end{figure}

To fully explore these capabilities for increasingly large networks, we again use the REGNN trained in Figure \ref{fig2} in random wireless networks of increasing size. Note that, as we increase the size of the networks, the density of the network remains constant so that the statistics of the channel conditions are the same. In Figure \ref{fig_full_test1}, we show the average sum-rate achieved by the REGNN over many random iterations for networks of increasing size $m^{\prime}$, where the geometric configurations generated using the fixed-density random geometric graph as done previously. We observe that, even as the network size increases, the same REGNN is able to outperform the heuristic methods. 

\begin{figure}
\centering
\includegraphics[height=.22\textheight, width = \linewidth]{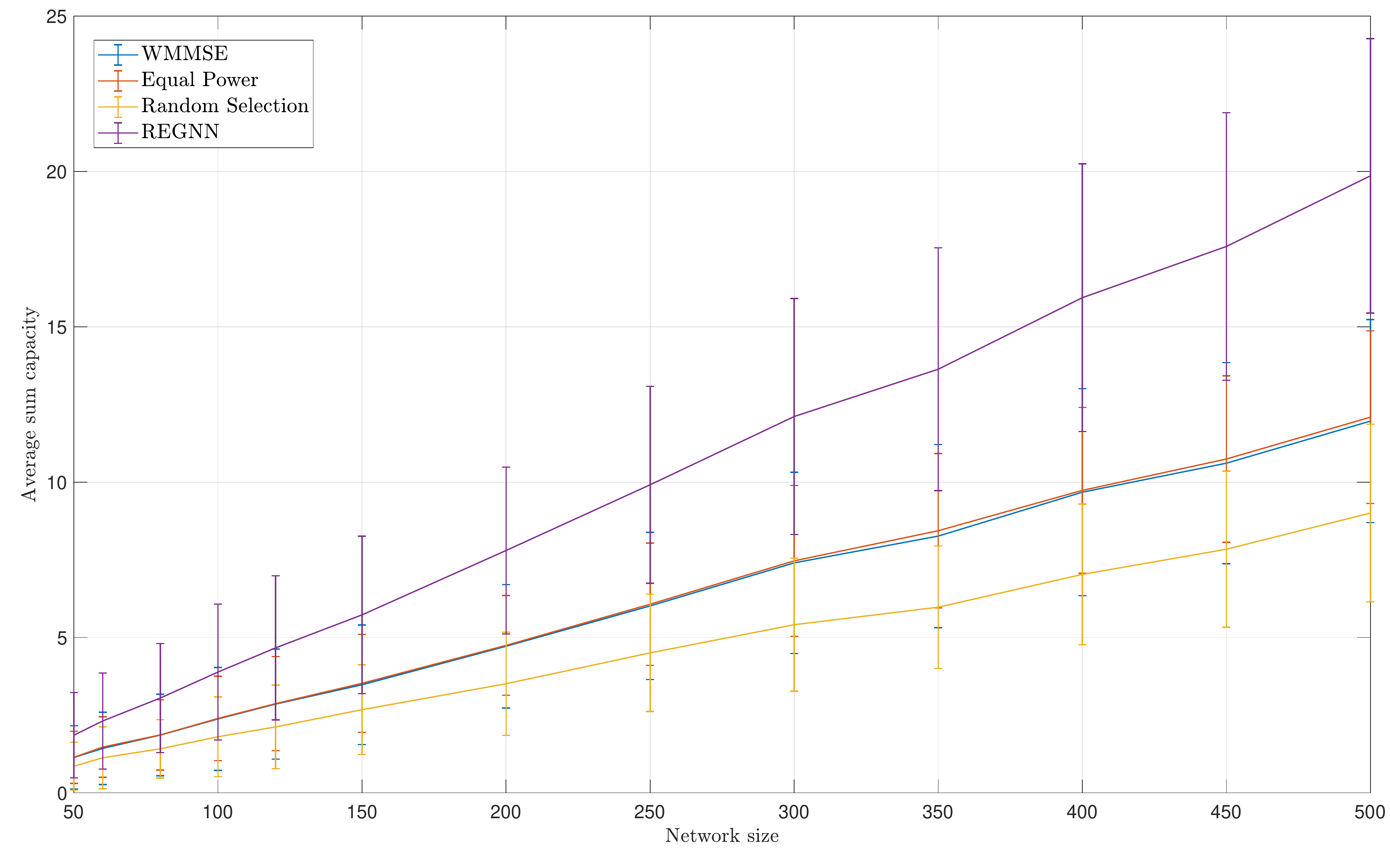}
\caption{Performance of REGNN trained in Figure \ref{fig2} in randomly drawn networks of varying size. From networks of size $m^{\prime} = 50$ to 500, the REGNN is able to outperform the heuristic methods.}
\label{fig_full_test1}
\end{figure}

As a final numerical study for the pairwise network, we compare the performance of the REGNN trained on a fixed network of size $m=50$ in new random networks of equivalent number of pairs but varying density. In these experiments, we draw random geometric graphs with some density factor $r$ by dropping transmitter $i$ uniformly at random at location $\bbt_i \in [- r^{-1} m \sqrt{m^{\prime}/m}, r^{-1} m \sqrt{m^{\prime}/m}]^2$, with its paired receiver at location $\bbr_i  \in [\bbt_i - m/4, \bbt_i + m/4]^2$ around its paired transmitter. In this manner, as the density factor $r$ increases, the physical space of the network gets smaller and thus more dense.  In Figure \ref{fig_full_test22}, we show the average sum-rate achieved by the REGNN over many random iterations for networks of increasing densities $r$. We observe that, for wireless networks of equal or less density than the one used for training, the REGNN has strong performance relative to the heuristics. However, as the networks more dense, the REGNN is unable to match the performance of WMMSE. This results follows from the fact that the statistics of data seen in training begins to vary more greatly from that seen in execution time as the networks increase in density. Indeed, as the transmitters become closer together, the path-loss component of the fading state decreases and the interference grows.

\begin{figure}
\centering
\includegraphics[height=.22\textheight, width = \linewidth]{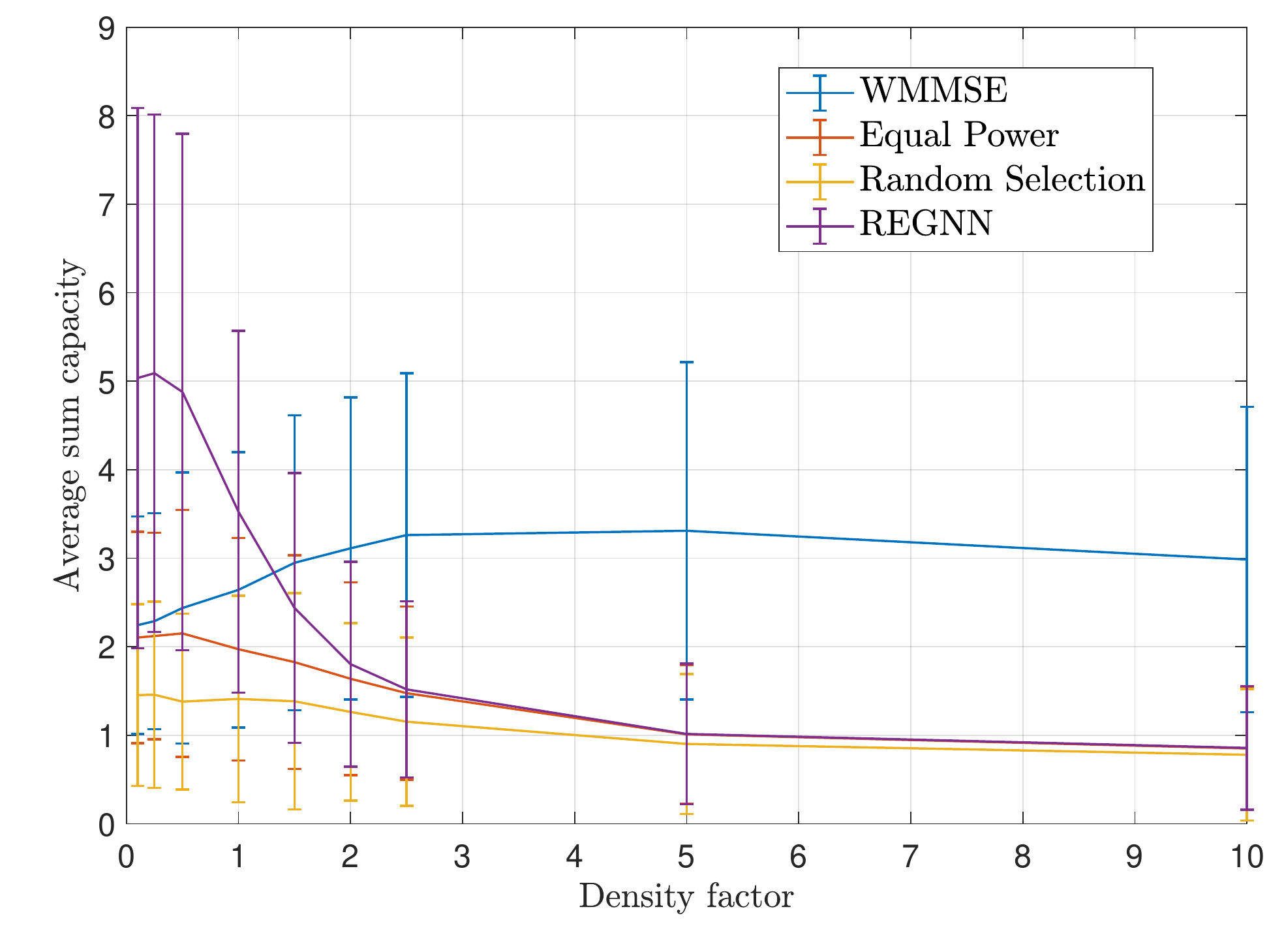}
\caption{Performance of REGNN trained in Figure \ref{fig2} in randomly drawn networks of varying densities from factors ranging from $r = 0.1$ to 10. As the density of the network increases, the REGNN is unable to match the performance of the WMMSE algorithm.}
\label{fig_full_test22}
\end{figure}

\subsection{Incorporating user demand}\label{sec_user_demand}

\begin{figure}
\centering
\begin{subfigure}[t]{\linewidth}
\centering
\includegraphics[width=\linewidth, height=.22\textheight]{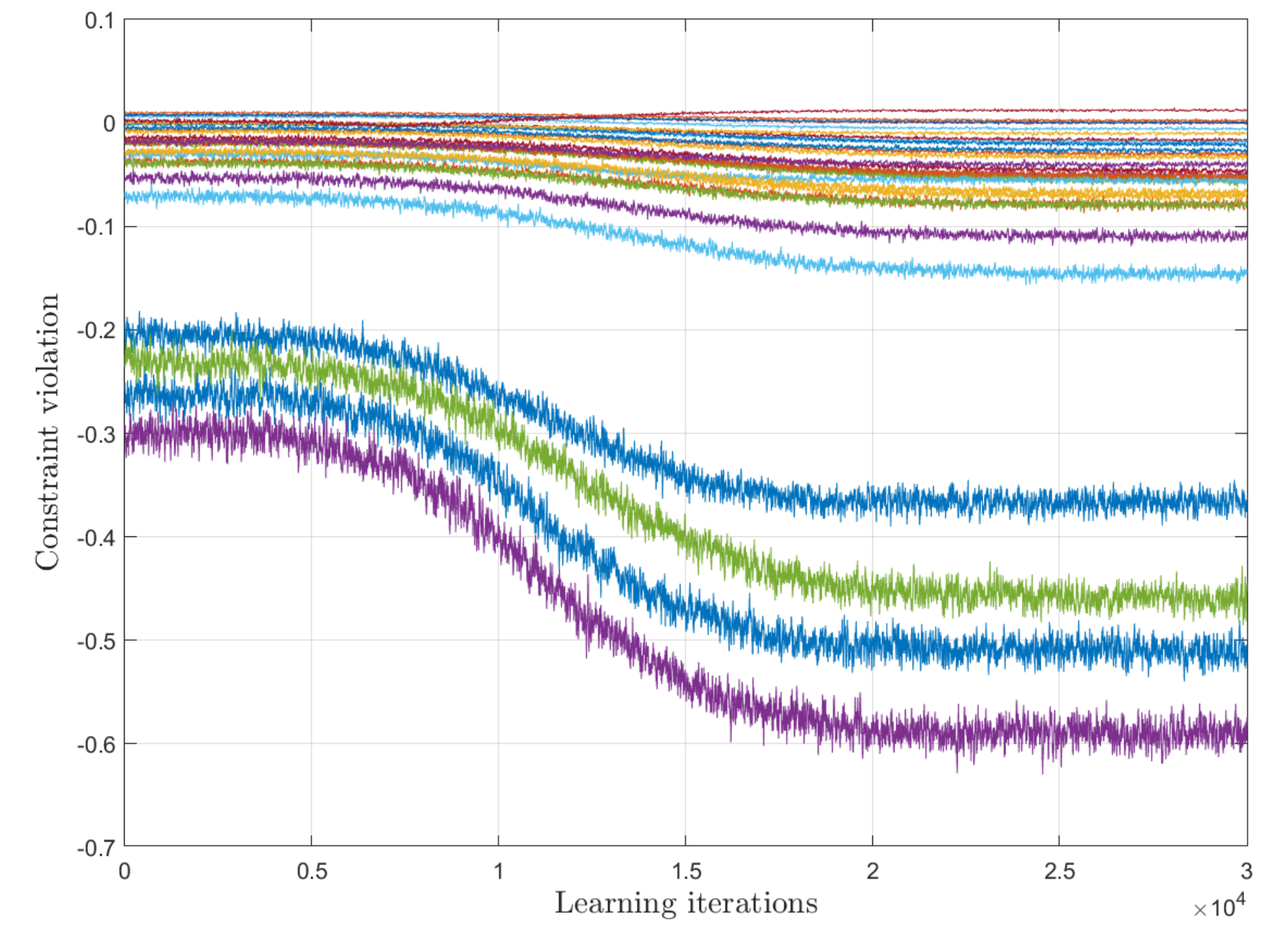}
\caption{}
\label{fig_data1a}
\end{subfigure}
\\
\begin{subfigure}[t]{\linewidth}
\centering
\includegraphics[width=\linewidth, height=.22\textheight,]{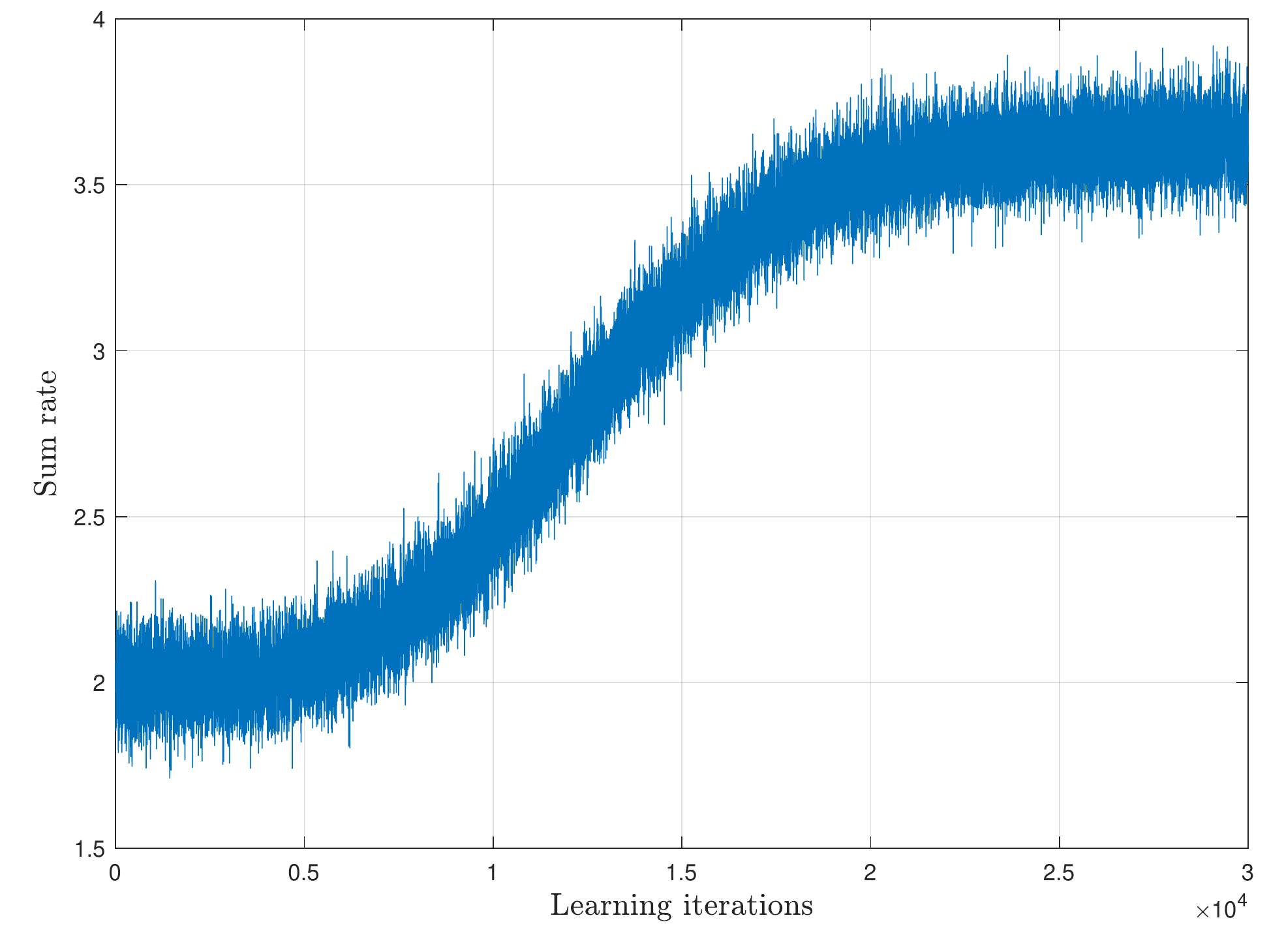}
\caption{}
\label{fig_data1b}
\end{subfigure} 
\caption{Convergence of training of an REGNN for the wireless sensor network problem with $m=30$ sensors/transmitters. In the top figure, we show the constraint violation for each of the transmitters converges to a feasible solution. In the bottom figure, we show the objective function converge to a local maximum.}
\label{fig_data1}
\end{figure}

We proceed to perform simulations an extension to the binary control problem in the paired ad-hoc network studied in the Section \ref{sec_adhoc}---namely binary power control with varying user demand. The problem was previously discussed in the second example of Section \ref{sec_examples}. For this setting, we assume that each transmitter additionally maintains a local state $x_i$ that reflects its current capacity demand, such as a packet arrival rate. Such a problem is highly relevant in, e.g. sensor networks and robotics applications. The resource allocation problem is closely related to that in \eqref{eq_power_control}, but with an additional constraint the ergodic capacity achieved by each transmitter must exceed the average collection rate of its associated sensor. We may write this problem as
\begin{align} \label{eq_power_control2}
   P^* & :=   \max_{\bbp(\bbH,\bbx),\bbr} \  \sum_{i=1}^m r_i,             \\
        &  \st            \         r_i      =  \E \Big[  \log \bigg(1 + \frac{|h_{ii}|^2 p_i(\bbH)}
              {1 + \sum_{j \neq i}| h_{ji}|^2 p_j(\bbH)}\bigg) \Big], \nonumber  \\
        &    \qquad                      \E \left[ \bbx\right] \leq \bbr, \quad 
                                        \bbp(\bbH,\bbx) \in  \{0, p_0\}^m.   \nonumber%
\end{align}

The problem in \eqref{eq_power_control2} contains an additional complexity in the ergodic constraint, which must be independently satisfied by each transmitter. The optimal resource allocation policy is which obtains sufficient capacity is achieved by each transceiver pair in expectation, while then maximizing the sum-rate achieved over the network. 

We perform the primal-dual learning method to train a REGNN in a system with $m=30$ transmitter/receiver pairs, who are placed randomly as in previous simulations. The collection rates $\bbeta$ are drawn from a exponential distribution with mean $0.05$. We train a REGNN with $L=10$ layers, each with $F_l = 1$ filters of length $K_l = 5$. In Figure \ref{fig_data1} we show the performance of the learning procedure. In Figure \ref{fig_data1a}, we show the constraint violation for all 30 sensors. with a negative value of $\E [ \bbx] - \bbr$ signifying a satisfaction of the capacity constraint, i.e. the sensor is transmitting data faster that it collects. While we may observe a wide variance in the capacities achieved by different sensors, a close examination of Figure \ref{fig_data1a} shows that all but 1 sensor achieves constraint satisfaction (constraint satisfaction can be seen with a negative constraint violation value). Likewise, in Figure \ref{fig_data1b}, we show the REGNN-based resource allocation policy achieve an overall performance converges to local optimum as the rate constraints are being satisfied.

\subsection{Multi-cell interference network}\label{sec_cellular}


\begin{figure}
\centering
\includegraphics[height=.22\textheight, width = \linewidth]{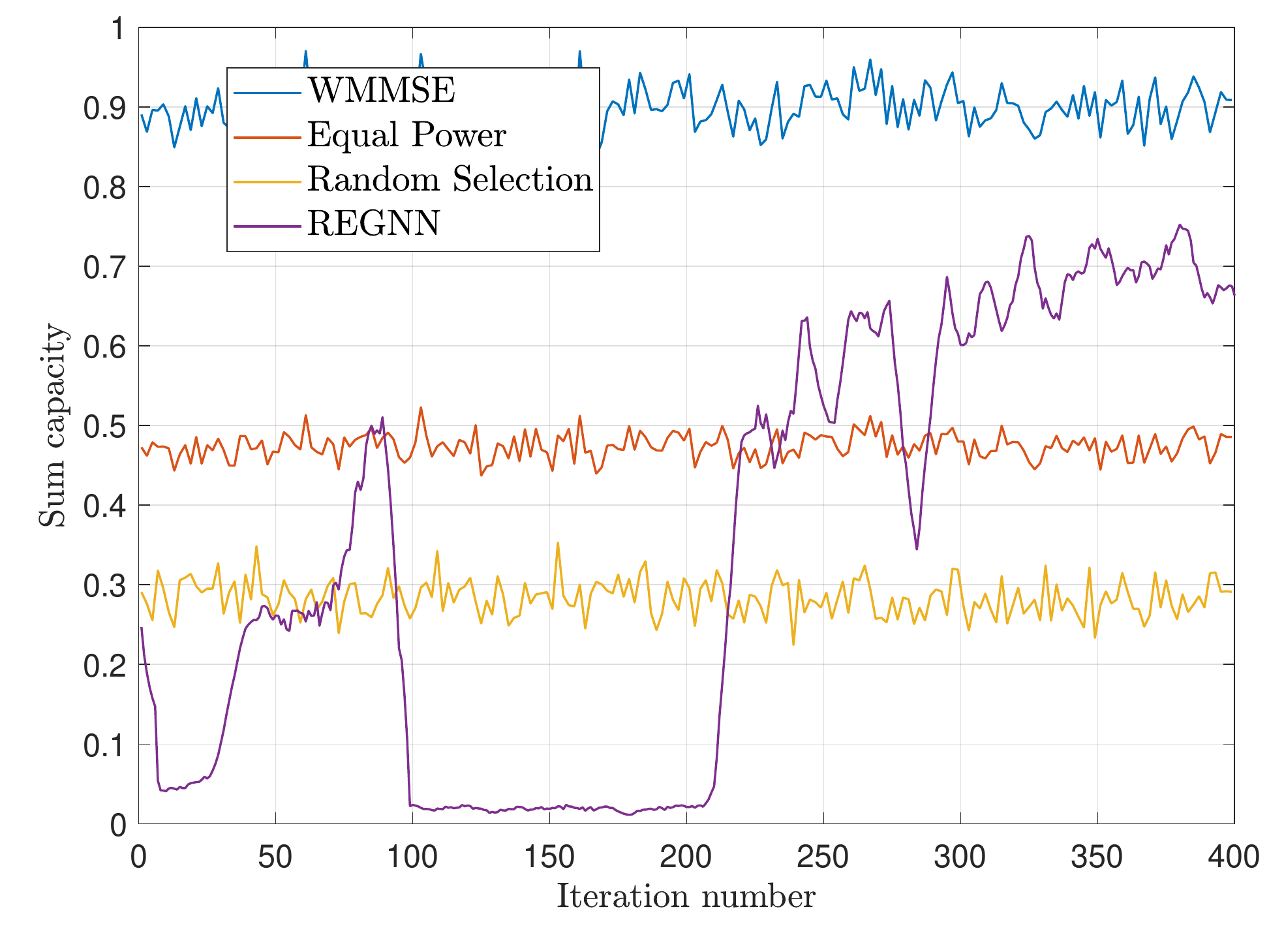}
\caption{Performance comparison during training of REGNN for multi-cell interference network with $m=50$ users and $n=5$ base stations. With $q=40$ parameters, the REGNN outperforms the model-free heuristics but does not quite meet the performance of WMMSE, which utilizes model information in its implementation. }
\label{fig_cell1}
\end{figure}

In this section, we consider a variation of the network architecture previously considered known as a multi-cell interference network. In contrast to the pair-wise setting, in the mutli-cell network there exist $n$ receivers---or cellular base stations---who service the transmissions of a total $m$ cellular users, which we assume are distributed evenly amongst the base stations. An example of such a multi-cell configuration is provided in Figure \ref{fig_systemb} for $n=20$ base stations, marked with large circles, covering a variety of cellular users, marked with smaller circles. The settings here is instructive not only in the real world practicality of its setting, but in its tendency to scale largely as the number of base stations or number of users grows. In Figure \ref{fig_cell1}, we show the performance obtained by the REGNN during training compared to the heuristic methods on a network of $n=5$ cells with a total of $m=50$ users. Here we see that the REGNN almost matches the performance of WMMSE without requiring model knowledge in its training or execution. As before, we study the transference properties as the number of cells grow. In Figure \ref{fig_full_test2}, we plot the performance of the REGNN trained in Figure \ref{fig_cell1} on networks of increasing size by increasing the number of cells. We observe that the performance of the REGNN scales well up to $n=30$ cells ($m=150$ users) and matches the performance of WMMSE without any model knowledge being utilized.


\begin{figure}
\centering 
\includegraphics[height=.22\textheight, width = \linewidth]{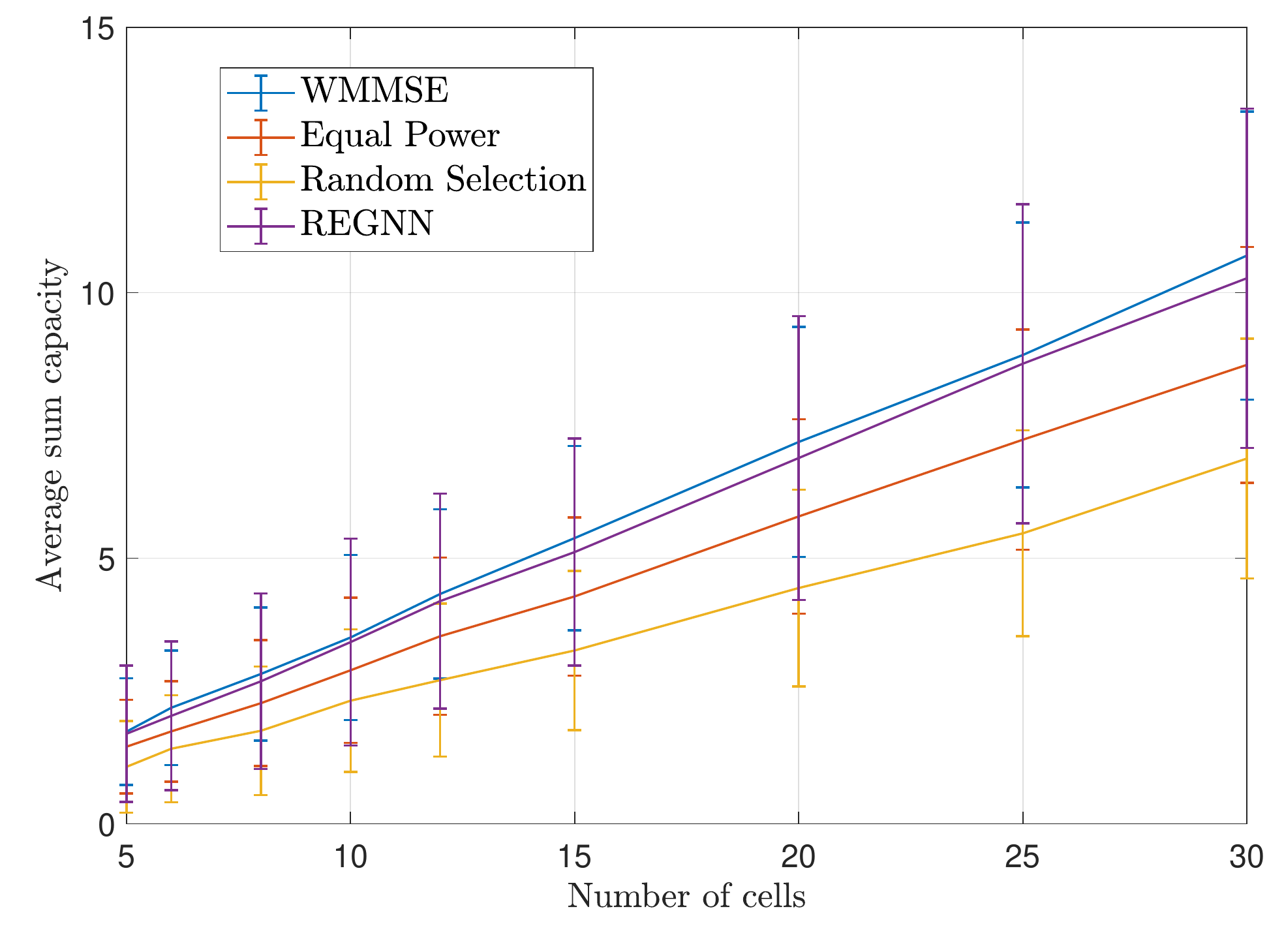}
\caption{Performance of REGNN trained in Figure \ref{fig_cell1} in randomly drawn multi-cell networks of varying size. From networks of size 5 to 50 cells, the REGNN matches the performance of the best performing heuristic method.}
\label{fig_full_test2}
\end{figure}

%
\section{Conclusion} \label{sec_conclusion}

We consider the problem of learning optimal resource allocation policies in wireless networks. The resource allocation problem takes the form of constrained statistical learning, which can be addressed by training a parameterization of the resource allocation policy using a model-free, primal-dual learning method. Fully connected neural networks are unsuitable parameterizations for large scale problems due to their prohibitively large parameter dimension. Given the randomly varying graph structure of fading channel states in a wireless network, we propose the use of random edge graph  neural networks (REGNNs) to parameterize the resource allocation policy. Such a neural network structure has significantly smaller parameter dimension that does not scale with the size of the wireless network. Moreover, we demonstrate that such policies are permutation equivariant, and can thus achieve similar performance on networks that are close to permutations of one another. We demonstrate in a series of numerical simulations how the REGNN is an effective parameterization for resource allocation policies for large scale wireless networks, both in learning strong performing policies and for such policies to transfer well to networks of larger size.

\urlstyle{same}
\bibliographystyle{IEEEtran}
\bibliography{wireless_learning}

\end{document}